\documentclass[12pt]{iopart}
\usepackage{graphicx}% Include figure files
\usepackage{iopams} % use  if AMS fonts are required

\begin{document}
%\pacs{\tt Not for circulation}

\title[Spectral properties of attractive bosons in a
ring]{Spectral properties of attractive bosons in a ring lattice including a single-site potential}

% \author{S M Cavaletto}
% \affiliation{Dipartimento di Fisica, Politecnico di Torino, Corso Duca degli Abruzzi 24, I-10129 Torino, Italy}
% \author{V Penna}
% \affiliation{Dipartimento di Fisica, Politecnico di Torino, Corso Duca degli Abruzzi 24, I-10129 Torino, Italy}
% \affiliation{CNISM, u.d.r. Politecnico di Torino, Corso Duca degli Abruzzi 24, I-10129 Torino, Italy}%
% \date{\today}

\bigskip
\author{ S M Cavaletto and V Penna}

\address{
Dipartimento di Fisica and Unit\`a C.N.I.S.M., Politecnico di Torino,
C.so Duca degli Abruzzi 24, I-10129 Torino, Italia}

\begin{abstract}
The ground-state properties of attractive bosons trapped in a ring
lattice including a single attractive potential well with an adjustable depth are investigated.
The energy spectrum is reconstructed both in the strong-interaction
limit and in the superfluid regime within the Bogoliubov picture. The analytical 
results thus obtained are compared with those found numerically
from the exact Hamiltonian, in order to identify the regions in 
the parameter space where this picture is effective. 
The single potential introduced is the simplest way to break the translational
symmetry and to observe, through a completely analytical approach, 
how the absence of symmetry affects the properties of the low-excited eigenstates
of the system. This model gives a first insight into the properties of systems
including more complex potentials.
\end{abstract}
% \maketitle

\section{Introduction}
\label{sez0}

Attractive bosons trapped in a one-dimensional (1D) periodic lattice \cite{JY}-\cite{OL} or in mesoscopic
arrays \cite{HC}-\cite{BPV3} have recently received considerable attention because they provide a natural
framework where Schr\"odinger-cat states \cite{Scat1}-\cite{Scat3} are in principle observable.  
Both the ground state and low-energy states of these systems, in fact, 
% have been shown to have the form of 
% consist of 
% have been shown to be 
have been shown to consist of superpositions of macroscopic spatially-localized quantum states
when the boson-boson interaction is sufficiently strong.
Meanwhile, a stimulating experimental work has made concrete the
realizability of lattices with a ring geometry \cite{AmOs} and the development of
optical-trapping schemes for engineering mesoscopic arrays \cite{AA1}, \cite{AA2}
in which Feshbach resonances \cite{Fes} ensure a full control of boson-boson interaction. 
$N$ bosons in a 1D periodic $M$-site lattice are well described within the Bose-Hubbard picture
by model Hamiltonian \cite{Hald}, \cite{Fisher}
\begin{equation}
H = \frac{\cal U}{2} \sum_j  (n_j^2 -n_j) - V -
T \sum_j ( a_{j+1}^+ a_j + a_{j+1} a^+_j) \, ,
%
%\sum_{<j \ell >} \, T_{j \ell } \,  a_j^+ a_\ell 
\label{BHH}
\end{equation}
where $n_i= a^+_i a_i$ ($i=1,..., M$), $a_{i+M} = a_i$  and $a_i$, $a^+_i$ obey the standard commutators 
$[a_m, a^+_i]= \delta_{mi}$. The boson tunneling among the lattice potential wells 
and boson-boson interactions are described by means of the hopping amplitude $T$ and parameter 
${\cal U} <0$ (${\cal U} >0$), respectively, in the attractive (repulsive) case.
In addition, term $V = \sum_i V_i \, n_i$ includes local potentials with well depths $V_i$ thus giving the
possibility to represent disordered lattices and/or trapping potentials. 

If $v_i =0$ for each $i$, the resulting system is homogeneous and features translation invariance. 
For ${\cal U} < 0$, the relevant zero-temperature scenario has been investigated in
\cite{BPV} and \cite{OL} showing how the resulting delocalized ground state exhibits three
characteristic regimes depending on the value of $\tau = T/(|{\cal U}| N)$.
In the strong-interaction (SI) regime, where $\tau < \tau_1 \simeq 1/4$, the ground state
is a Schr\"odinger cat well represented by a super-position of $M$ coherent
states of algebra su($M$), each one describing the strong localization of bosons around 
a given lattice site.
The opposite weak-interaction regime is the superfluid (SF) one, defined in the open
interval $\tau > \tau_2$ $ =1/[2M \sin^2 (\pi /M)]$. Its ground state reduces to a single
su($M$) coherent state describing the uniform distribution of bosons in the lattice and,
thus, their complete delocalization.
Last, for $\tau_1 <\tau < \tau_2$, the solitonlike regime features
a ground state which again is a superposition of $M$ localized states. The latter, however,
exhibit an intermediate character: their localization peaks describe boson distributions involving
a significant number of lattice sites. Peaks become sharper and sharper when $\tau \to \tau_1$.

This almost ideal scenario, where translation invariance
combined with the fact that ${\cal U } < 0$ enable Schr\"odinger-cat states
to appear, breaks up as soon as $V \ne 0$.
In this paper we investigate the low-energy properties of model (\ref{BHH})
and analyze, in particular, the crucial role played by a localized perturbation, a single
potential well, in modifying the structure of low-energy states and the relevant spectrum.
The interest for this model is supported by various motivations.
First, the introduction of a local potential well is the simplest 
possible way to introduce a disturbance in a perfectly symmetric ring lattice
characterized by translation invariance. Potential $V$
in $H$  will contribute with a single term at some site $j$
and $V_i = 0$ for any $i \ne j$. 
Moreover, this naive model preludes to a very realistic situation. In fact, the presence
of lattice defects --representable in terms of extremely weak local potentials-- should
be viewed as an intrinsic, essentially uncontrollable, ingredient of the experimental setup. 
If ${\cal U }>0$, their perturbative character does not affect large-scale phenomena such as
the formation of Mott and SF states and thus defects can be ignored. Conversely, for ${\cal U }<0$,
a single, even vanishingly small, defect is able to break the system symmetry therefore preventing the
formation of Schr\"odinger cats.
Finally, the model with a single-site potential has the non secondary advantage to allow one 
the implementation of standard approximation schemes and a fully analytic study of the
Hamiltonian and of its energy spectrum.
%
%In general, the case of an attractive bosonic ring with a single-site potential
%preludes to real systems where complex behaviors are known to take place in the
%presence of a random distribution of potential depths. In passing, we note that
%while disordered lattices with repulsive bosons have been widely explored
%so far, the case with attractive bosons has received no or scarce attention. 

We show that, in the SI regime
and in the presence of a single potential well (placed, for example, at $j=0$)
with depth $V_0$ even arbitrarily small, the localization effect of bosons enables
us to operate a remarkable simplification of $H$. The latter reduces to a pure-hopping model
including a new effective local potential whose depth is proportional to the total boson number
(and therefore is much larger than $V_0$). 
%
%The resulting Hamiltonian can be separated in two independent parts related to two new sets
%of bosonic modes (each one containing essentially $M/2$ modes) one of which is totally independent
%from the effective local potential. These take $H$ into a diagonal form. 
%
Such a new form is particularly useful since, with essentially no analytic work, $H$ can
be separated in two commuting sub-Hamiltonians one of which is intrinsically diagonal.
The most interesting feature, however, concerns the SP (single-particle) energy spectrum
obtained from the complete diagonalization of $H$. Depending on the choice of parameters
$T$, $\cal U$ and $V_0$, the spectrum
exhibits a structure characterized either by well-visible energy doublets
or by an almost uniform distribution of SP energies. The notable exception in the latter case
is the lowest SP energy showing, for $\tau \to 0$, an unexpected diverging behavior to 
arbitrarily large negative values. This feature has been already observed in the study of the
SP spectrum of bosonic comb lattices \cite{parma} and of their unusual zero-temperature properties. 

Then we investigate the SF regime by adopting 
the standard Bogoliubov picture to recast our $V\ne 0$ model into a
more convenient form. In this case, however, a rather hard mathematical work is necessary
to diagonalize $H$. To this end, after recognizing that $H$ can be expressed as a linear
combination of the operators belonging to $M$ independent su(1,1)-like algebras,
a remarkable help in the diagonalization process comes from exploiting the transformation
properties of such algebras.
Also in this case, the definition of two independent sets of new bosonic modes
makes $H$ separable into two commuting sub-Hamiltonians
% involving two independent sets of new bosonic modes 
one of which
exhibits the characteristic SP energies distinguishing the solution of SF models within the
Bogoliubov diagonalization scheme.
The SP energies of the 
other sub-Hamiltonian (this is written in terms of $V_0$-dependent modes)
are found to represent small deviations of
Bogoliubov-like SP energies. The SP spectrum is thus characterized by energy doublets.
Both in the SI and in the SF case, we determine the analytic form of weakly excited states.

Sections II and III are devoted to the study of the SI regime and the SF regime, respectively.
In both sections the validity of analytic results concerning the ground state and the first few weakly-excited
energy levels are compared with numerical results. Section IV is devoted to final comments.

%%%%%%%%%%%%%%%%%%%%%%%
% It is worth noting that the concrete realization of this simple model seems to be within the reach 
% of current techniques. In the recent years, the exceptional progress in optical-trapping techniques
% has allowed to engineer complex disordered lattices involving many wells, superlattices or arrays formed by a few wells.
% Paradoxically, this task appears to be simpler than the realization of a ring with single well. 
%%%%which however does not seem to be beyond the reach of current trapping techniques. 
% In this respect an encouraging example is the realization of toroidal traps \cite{Ryu} with a persistent flow.
% This system has raised a lot of interest because gives the possibility to create a bosonic Josephson junction.
% The latter could be achieved by intersecting the toroidal domain with a transverse laser beam generating a
% potential barrier. This proposal suggests that the our model (involving a well instead of a barrier) could represent
% more than a simple but interesting toy model.
%
%%%%%%%%%%%%%%%%%%%%%%%%%%%%%%%%%%%%%%%%%%%%%%%%%%%%%%%%%%%%%%%%%%%%%%%%%%%%%%%%%%%%%%%%%%%%%%%%%%%%%%%%%%%%%%%
\section{ Low-energy states in the SI regime}
\label{sez2}

Low-energy states of Hamiltonian (\ref{BHH}) with $V=0$ and ${\cal U} < 0$ consist of
a superposition of su($M$) coherent states, each one involving a boson distribution strongly
localized around a different site of the lattice. Such coherent states are defined by
$$
|\xi \rangle \, := \frac{1}{\sqrt { N!}} 
\left ( \sum^{M-1}_{i=0} \xi_i  a^+_i \right )^N |0\rangle\, , \quad
\langle \eta| \xi \rangle = \left ( \sum^{M-1}_{m=0} \eta^*_m \xi_m \right)^N
$$
where $N$ is the total boson number, state $|0\rangle$ is such that $a_i |0\rangle = 0$ 
for each $i$, and
the second equation defines the scalar product of two generic states. 
The normalization of $\langle \xi | \xi \rangle$
is thus ensured by $\sum_m \xi^*_m \xi_m =1$.
Based on the previuos definition, 
the explicit form of low-energy states in terms of localized states $|\xi (j) \rangle$
was found to be \cite{BPV}
\begin{equation}
|E_k \rangle = \sum^{M-1}_{j=0} \, \frac{e^{i{\tilde k}j}}{\sqrt M} \, |\xi (j) \rangle
\, , \qquad
|\xi_i (j)| << |\xi_j (j)| \simeq 1
%
%|\xi (j) \rangle \, := \frac{e^{i{\tilde k}j}}{\sqrt { N!}} 
%\left ( \sum_i \xi_i(j) a^+_i \right )^N |0\rangle
\label{GS1}
\end{equation}
in which index ${\tilde k} = 2\pi k/M$ 
with $k \in [0, M-1]$ essentially represents the eigenvalue of the quasi-momentum operator
generating lattice translations. The ground state corresponds to the case $k=0$.
The localization of bosons at site $j$ is embodied in
inequality $|\xi_i (j)| << |\xi_j (j)|$ of formula (\ref{GS1}), 
the quantity $|\xi_\ell (j)|^2$ representing the fraction
of population at site $\ell$ according to state $|\xi (j) \rangle$.
This feature becomes evident by reminding that
$\langle \xi (j)| a^+_\ell a_\ell |\xi (j) \rangle = N |\xi_\ell (j)|^2$ \cite{BP}.
In addition, because states $|\xi (j) \rangle$ 
can be equipped with the property $\langle \xi (i)  |\xi (m) \rangle = \delta_{im}$,
one can prove \cite{BP} that $\langle E_k | a^+_i a_i |E_k \rangle = N/M$ whose site independence
confirms that $|E_k \rangle $ are delocalized states. 
For $\tau < 1/4$, states $|E_k \rangle$ have been shown \cite{BPV} 
to provide a quite satisfactory approximation
of the true energy eigensates whose exact form can be found only numerically.

In the ideal lattice ($V=0$) the model features translation symmetry which is
responsible for the super-position of equal-weight localized states in state (\ref{GS1}).
In the classical limit, this symmetry is broken:
only one of components $|\xi (j) \rangle$ survives giving rise to the
exponential localization \cite{ILLU} that is known to distinguish the maximally excited state
of model (\ref{BHH}) with ${\cal U} > 0$.
The presence of local potential $V = V_0 n_0 $ ($V_0 >0$) in our quantum model
\begin{equation}
H  = -\frac{U}{2} \sum_i n_i (n_i-1)- V_0 n_{0} - T \sum_i ( a^+_{i+1} a_i  + \mathrm{H.C.} )
\label{HV}
\end{equation}
with ${\cal U} = -U <0$
also breaks this symmetry suggesting that only one among the $M$ components of
$|E_0 \rangle$ is expected to survive.
If one assumes that the most part of the population is
placed at site $j=0$ owing to the presence of the attractive well, 
then Hamiltonian (\ref{HV}) can be taken into a new 
approximate form
whose diagonalization appears to be rather simple. The approximation we
effect essentially coincides with the Bogoliubov scheme. Observing that
$N= \sum_i \, n_i \, \Rightarrow \, n_{0} = N - \delta N$ where $\delta N = {\sum_{i\ne 0}} n_i$
and $n_i << n_0 \simeq N$ for $i \ne 0$, then
$$
\sum_i n_i^2 
=  {\sum_i}^* n_i^2 + N^2 + (\delta N )^2 - 2N\, \delta N \simeq -N^2 +2N\, n_0
$$
in which terms $n_i n_m$ with $i,m \ne 0$ have been suppressed. Model (\ref{HV}) thus reduces to
\begin{equation}
H  \simeq 
\frac{U}{2} N(N+1) - w n_0 - T {\sum}_j ( a^+_{i+1} a_i + \mathrm{H.C.})
\label{HVo}
\end{equation}
in which the assumed localization at $j=0$ has the dramatic effect to involve
a much deeper (effective) well with depth $w= U N+V_0$ together with the disappearence of
nonlinear interaction terms $n_i(n_i -1)$. 
Therefore, even if the initial well is a simple perturbation
where $V_0$ could be vanishingly small, the depth of the resulting effective well
can be really large since it depends on $UN$.
The role of attractive interaction ${\cal U} =$ $ -U<0$ is thus to reduce the initial model
to a pure-hopping model with a deeper effective well. 

%%%%%%%%%%%%%%%%%%%%%%%%%%%%%%%%%%%%%%%%%%%%%%%%%%%%%%%%%%%%%%%%%%%%%%
\subsection{Diagonalization}

After assuming Hamiltonian (\ref{HVo}) as the reference model in the SI regime,
its diagonalization is performed by resorting to the momentum-mode picture
\begin{equation}
a_j = 
M^{-\frac{1}{2}} \sum_k  b_k\, {e^{i{\tilde k}j } } 
\,\, \Leftrightarrow \,\,
b_k = 
M^{-\frac{1}{2}} \sum_k  a_j\, {e^{-i{\tilde k}j }} \,  ,
\label{fab}
\end{equation}
with ${\tilde k}= 2\pi k/M$ and  $j, k \in [0, M-1]$, where
$a_{j+M} \equiv a_j $ and $b_{k}\equiv b_{k+M}$ owing to the periodic boundary
conditions of the lattice. This gives
\begin{equation}
H_w =   C_N - w\, a^+_0 a_0 - 2T {\sum}_k \, c_k b^+_k b_{k}\, ,
\label{HB2}
\end{equation}
where $a_0 = \sum_k b_k /{\sqrt M}$, $C_N={U}N(N+1)/2$ and
$c_k = \cos ({\tilde k})$. To achieve the diagonal form of $H_w$
it is particularly advantageous to define the new operators
\begin{equation}
f_k = (b_k -b_{-k})/{\sqrt 2} \, , \quad F_k = (b_k +b_{-k})/{\sqrt 2}\, ,
\label{newop}
\end{equation}
and $F_0 = b_0$, $f_0 = 0$. In case $M$ is even, the further operator
$F_{M/2} = b_{M/2}$ must be included while $f_{M/2}= 0$. The range of index $k$ is such that 
\begin{equation}
1 \le k \le S= (M-1)/2\, ,\quad 1 \le k \le S= (M-2)/2\, ,
\label{Sran}
\end{equation}
if $M$ is odd or even, respectively. Note that such operators satisfy 
the usual bosonic commutators $[f_n ,f^+_h  ] = \delta_{nh} = [F_n ,F^+_h  ]$.
Then $H_w$ becomes
\begin{equation}
H_w = 
C_N - w  a^+_0 a_0 - 2T \sum^S_{k=1} c_k f^+_k f_{k}
- 2T \sum^K_{k=0} c_k F^+_k F_{k} 
\label{HB2fF}
\end{equation}
with
$$
a_0 = {\sum}_k { b_k }/{\sqrt M } = {\sum}^K_{k=0}  r_k { F_k }/{\sqrt M } \, ,
%
%\left [ {\sqrt {2}} \sum_k  F_k \right ]
$$
$r_0 = r_{M/2} = 1$, $r_k = {\sqrt {2}}$, and the range of $k$ given by
\begin{equation}
0 \le k \le K= (M-1)/2\, ,\quad 0 \le k \le K= M/2\, ,
\label{Kran}
\end{equation}
when $M$ is odd or even, respectively. 
Hamiltonian (\ref{HB2fF}) is thus formed by two commuting parts one of which, 
$H_f = \frac{U}{2} N(N+1) - 2T \sum^S_{k=1} \, c_k f^+_k f_{k}$, is diagonal.
The remaining part,
$$
H_F = - w\, a^+_0 a_0 - 2T \sum^K_{k=0} \, c_k F^+_k F_{k} = -\sum^K_{h,k=0} \, L_{kh} F^+_k F_{h}\, ,
$$
with $L_{kh} = w\, r_k r_h/M+2T c_k \delta_{kh}$, can be diagonalized in a rather direct way. In fact, since 
$L_{kh}$ are elements of an $M\times M$ real and symmetric matrix, there exists an orthogonal transformation
of elements $B_{pk}$ such that $\sum_p B_{pk}B_{ph}\, =\, \delta_{kh}$, $\sum_k B_{pk}B_{qk}\, =\, \delta_{pq}$ and, in particular,
\begin{equation}
{\sum}_{kh}B_{pk}L_{kh}B_{qh}\, =\, \lambda_{p}\delta_{pq} \, .
\label{orthogdiag}
\end{equation}
The latter entails $L_{kh}\, =\, \sum_{pq}B_{pk}B_{qh}\lambda_{p}\delta_{pq}$. As a result, by introducing the new
bosonic creation and annihilation operators $D_p\, =\, \sum_k B_{pk}F_k$ and $D_q^+\,=\, \sum_h B_{qh}F_h^+$, 
satisfying standard bosonic commutators due to the orthogonality of $B_{pk}$, 
one gets the diagonal form
$$
H_F\, =\,  -{\sum}_q\lambda_q D^+_q D_q\, .
$$
%%implying, of course, $i\hbar {\dot D}_p = [D_p, H_F] =-\lambda_p D_p$. 
% 
By using the previous definition of $L_{kh}$ together with the orthogonality relations
for $B_{pk}$, equation (\ref{orthogdiag}) becomes
$\sum_{kh}B_{pk}(w r_k r_h/M+2T c_k \delta_{kh})B_{qh} =\lambda_{p} \sum_k B_{pk}B_{qk}$
and thus
\begin{equation}
{\sum}_{kh}B_{pk}[w\, r_k r_h/M-(\lambda_p\, -\, 2T\, c_k) \delta_{kh}]B_{qh}\, =\, 0\, .
\label{orthogdiag2}
\end{equation}
In order to satisfy the latter equation, we define $A(p)\,=\,\sum_h r_h  B_{ph}$ and 
impose  $(\lambda_p - 2Tc_k)\, B_{pk} = {w r_k} \, A(p)/{M} $ for each $p$ obtaining
\begin{equation}
%(2Tc_k -\lambda_p)\, B_{pk} = -\frac{w r_p}{M} \, A(p)\, \,
% 
B_{pk} = \frac{wr_k}{M} \,\frac{ A(p)}{\lambda_p - 2Tc_k} \, .
\label{B}
\end{equation}
This definition, inserted in the orthogonality relation $\sum_k B_{qk} B_{qk}\,=\,1$, gives
\begin{equation}
|A(p)|^2 =  
\frac{ M^2}{w^2} \left [ \, {\sum}_k\,  \, \frac{r_k^2 }{(\lambda_q - 2Tc_k)^2} \right ]^{-1} \, ,
\label{Ap}
\end{equation}
which enables one to fix parameter $A(p)$.
By multiplying both sides of (\ref{B}) times $r_k$ and summing over $k$, one easily derives the 
crucial formula
%
% $$ {\sum}^K_{k=0} r_k B_{pk} = \frac{w}{M} {\sum}^K_{k=0} \,
% \frac{r_k^2 A(p)}{\lambda_p - 2Tc_k} \, \, $$
%
\begin{equation}
1\, = \frac{w}{M}\, {\sum}^K_{k=0} \, \frac{r^2_k }{\lambda_p- 2Tc_k} \, ,
\label{EV}
\end{equation}
determining eigenvalues $\lambda_p$ and thus the $H_F$ spectrum.
To conclude, the total (diagonal) Hamiltonian reads
\begin{equation}
H_w = C_N 
-2T {\sum}^S_{k=1} \, c_k f^+_k f_k -{\sum}^K_{q=0} \, \lambda_q D^+_q D_q \, .
\label{DH1}
\end{equation}
By observing that the vacuum state of operators $f_k$ and $D_p=\sum_k {B}_{pk} \, F_k$ 
(these are  linear combinations of $b_k$ and $b_{-k}$)
coincides with that of modes $b_k$ defined by 
$b_k |0\rangle = 0$ for each $k$ ($|0\rangle \equiv |0, 0, \, ..\, 0\rangle$),
the Fock states relevant to $f_k$ and $D_p$ are found to be
\begin{equation}
|{\vec \ell}, {\vec m} \rangle
%%%|\ell_1, \, ...\, \ell_S\rangle | m_0, m_1, \, ...\, m_K \rangle=
%
= \prod^S_{k=1} \, \frac{ (f^+_k)^{\ell_k} }{\sqrt {\ell_k !}} \, 
\prod^K_{p=0} \, \frac{ (D^+_p)^{m_p} }{\sqrt {m_p !}} \, |0\rangle
\label{fock}
\end{equation}
satisfying
$D^+_q D_q \,|{\vec \ell}, {\vec m} \rangle = m_q \,|{\vec \ell}, {\vec m} \rangle$
and
$f^+_k f_k \,|{\vec \ell}, {\vec m} \rangle = \ell_k \,|{\vec \ell}, {\vec m} \rangle$.
The relevant eigenvalue equation 
$H_w \, |{\vec \ell}, {\vec m} \rangle = E({\vec \ell}, {\vec m}) |{\vec \ell}, {\vec m} \rangle$
features energy eigenvalues
\begin{equation}
E({\vec \ell}, {\vec m})=
C_N -2T {\sum}^S_{k=1} \, c_k \ell_k -{\sum}^K_{q = 0} \, \lambda_q m_q\, .
\label{ener}
\end{equation}
In particular, the ground state, in which all bosons possess 
the lowest SP energy  $-\lambda_0$ and therefore
$\ell_k = m_q =0$, $m_0 =N$ (the assumption that 
$\lambda_0 > \lambda_p, \, 2Tc_k$, $\forall\, p\ne 0$ and $\forall\, k$ will be
discussed in the next section), is given by 
\begin{equation}
|\mathrm{GS} \rangle =
%|0, 0, \, ...,\, \rangle | N, 0, \, ...\, 0 \rangle=
\frac{ (D^+_0)^{N} }{\sqrt {N !}} \, |0 \rangle=
\frac{ 1 }{\sqrt {N !}} \left (
{\sum}^K_{k=0} {B}_{0k} \, F^+_k \right )^{N}\, |0 \rangle
\label{GS}
\end{equation}
which exhibits the form of a su($M$) coherent state. This feature pertains as well to
excited states such as ${ (D^+_p)^{N} }\, |0 \rangle /{\sqrt {N !}} $ and
%|0, \, 0 ...\, N ,\, 0 ...\rangle | 0, 0, \, ...\, 0 \rangle=
${ (f^+_k)^{N} }/{\sqrt {N !}} \, |0\rangle $ characterized by the fact that all bosons
condensate in a specific single-particle energy  $-\lambda_p$ and $-2T c_k$, respectively.
In the second case, however, bosons are distributed only between momentum states $+k$ and $-k$,
and the total momentum turns out to be zero since 
$\langle b^+_k b_k \rangle$ = $\langle b^+_{-k} b_{-k} \rangle=N/2$.
Any other excited state is represented by state (\ref{fock}).

%%%%%%%%%%%%%%%%%%%%%%%%%%%%%%%%%%%%%%%%%%%%%%%%%%%%%%%%%%%%%%%%%%%%%%%%%%%%%

\subsection{Spectrum of $H_w$}
\label{22}

Exact SP  energies $-\lambda_p$ can be obtained numerically from formula (\ref{EV}).
Nevertheless, analytic approximate solutions of this equation can be found in two
limiting cases which give interesting information on the spectrum structure.
In order to calculate eigenvalues $\lambda_p$ 
%representing the single-particle spectrum of Hamiltonian $H_F$, 
we rewrite formula (\ref{EV}) as
\begin{equation}
\frac{2T M}{w}=  {\sum}^K_{k=0}  \frac{r^2_k }{ \mu -  c_k  }
\equiv {\sum}^{M-1}_{k=0}  \frac{1 }{ \mu -  c_k  } \, , 
\label{SPS}
\end{equation}
where 
$$
\mu = {\lambda}/({2T})\, , \quad
c_k = \cos (2\pi k/M)\, ,
$$ 
and one should remind that $r^2_0 = r^2_{M/2} =1$ while $r^2_k = 2$, $\forall k \ne 0, M/2$.
The number of solutions depends on $M$: equation (\ref{SPS}) gives $M/2 +1$ solutions for $M$ even and 
$(M+1)/2$ with $M$ odd (see appendix \ref{A2}).
Interestingly, series (\ref{SPS})
can be written in terms of either hyperbolic or trigonometric functions
depending on the fact that $|\mu | >1$ or $|\mu | <1$, respectively
(see, for example, \cite{Hans}). Then, after introducing parametrizations 
$\mu = \, {\rm ch} y$ and $\mu =  \,\cos y$,
involving identities (\ref{hyperF}) and (\ref{trigF}), respectively,
we obtain the alternative forms of equation (\ref{SPS})
\begin{equation}
({2T}/{w}) \, {\rm sh} y =  {\rm cth} ( {My}/{2} )
\, ,
\label{SPS1}
\end{equation}
\begin{equation}
({2T}/{w}) \sin y = - {\rm ctg} ( {My}/{2} )
\, .
\label{SPS2}
\end{equation}
Equation (\ref{SPS1}) is able to supply only one
solution, as follows from the comparison of functions 
$z= 2T\, {\rm sh} y / {w}$ and $z=  {\rm cth} ( {My}/{2} )$
in the $yz$ plane.
In the two cases $T/w >> M/4$ and $T/w << M/4$ the corresponding curves $z(y)$ intersect
at low and large values, respectively, of $y$. In this limits, one easily finds that
$$
y\simeq {1}/{\sqrt {(TM/w) - M^2/8} }\, , \qquad y\simeq {\rm arcsh }(w/2T)\, ,
$$
giving the SP energies
\begin{equation}
\mu \simeq 1 + \frac{w}{2TM} \, ,\qquad \mu \simeq \sqrt {1+ \frac{w^2}{4T^2} }\, ,
\label{Ssing}
\end{equation}
for $T/w >> M/4$ and $T/w << M/4$, respectively. According to Hamiltonian (\ref{HB2fF}) 
the effective well depth is $w = V_0 +UN$ where, even if $V_0$ is small,
$UN$ and thus $w$ are large. Since we are considering the SI regime in which $T/UN <1$,
the second case where $\mu ={\lambda}/{2T} \simeq \sqrt {1+ {w^2}/{4T^2} }$ is the interesting one.

Concerning equation (\ref{SPS2}), the two cases $T/w >> M/4$ and $T/w << M/4$ once more allow one
to distinguish the significant regimes of this equation and the ensuing solutions.
Approximate solutions of equation (\ref{SPS2}) are found by substituting
in this equation $y = {y}_k  + \epsilon_k$ with ${y}_k = 2\pi k/M$ if $T/w >> M/4$
and $y = {\bar y}_k  + \epsilon_k$ with ${\bar y}_k ={y}_k +\pi/M$ if $T/w << M/4$. 
Parameters $\epsilon_k$ are such that $|\epsilon_k| << |{y}_k|, \, |{\bar y}_k|$.
The latter assumption and the ensuing calculation are discussed in appendix \ref{A3}.
We obtain, for $T/w >> M/4$,
\begin{equation}
\mu_k = \frac{\lambda_k }{2T} 
= \cos ( {y}_k + \epsilon_k) \simeq \, \cos ( {y}_k) \, +  \frac{w}{T M}  \, ,
%({\rm for} \, 2T/w >> M/2)
\label{sp1}
\end{equation}
and, for $T/w << M/4$,
\begin{equation}
\mu_k = \frac{\lambda_k }{2T} 
= \cos ( {\bar y}_k + \epsilon_k) \simeq
\, \cos ( {\bar y}_k) \, +  \frac{4T}{wM}\, \sin^2 ({\bar y}_k) \, .
\label{sp2}
\end{equation}
In both regimes, the set of SP energies $-\lambda_k$ relevant to the $D$-mode component 
of Hamiltonian (\ref{DH1}) is completed by the energies $-2T \cos (y_k)$ of the $f$-mode
component.

\begin{figure}
\begin{center}
\includegraphics[width=6.0cm]{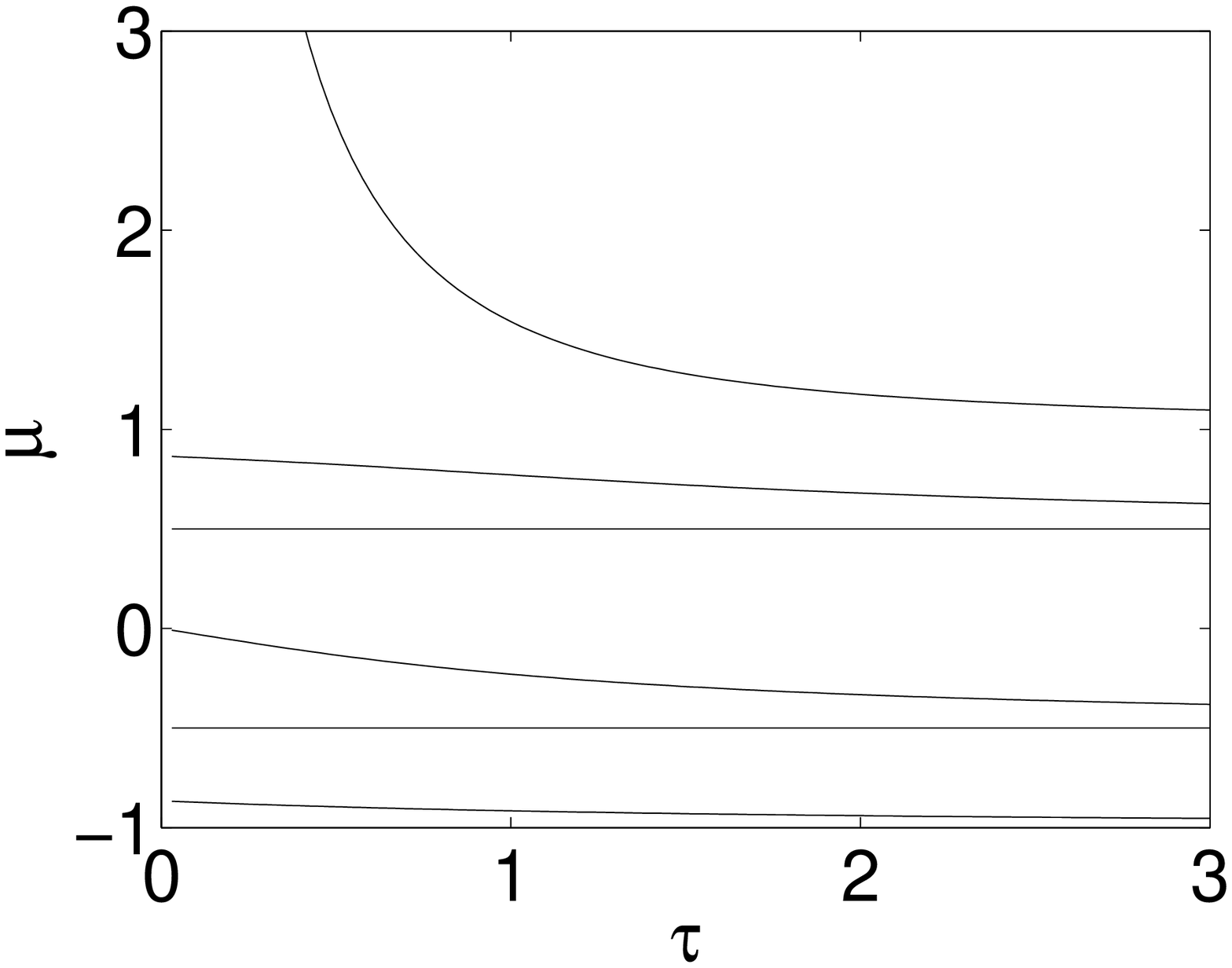}
\includegraphics[width=6.0cm]{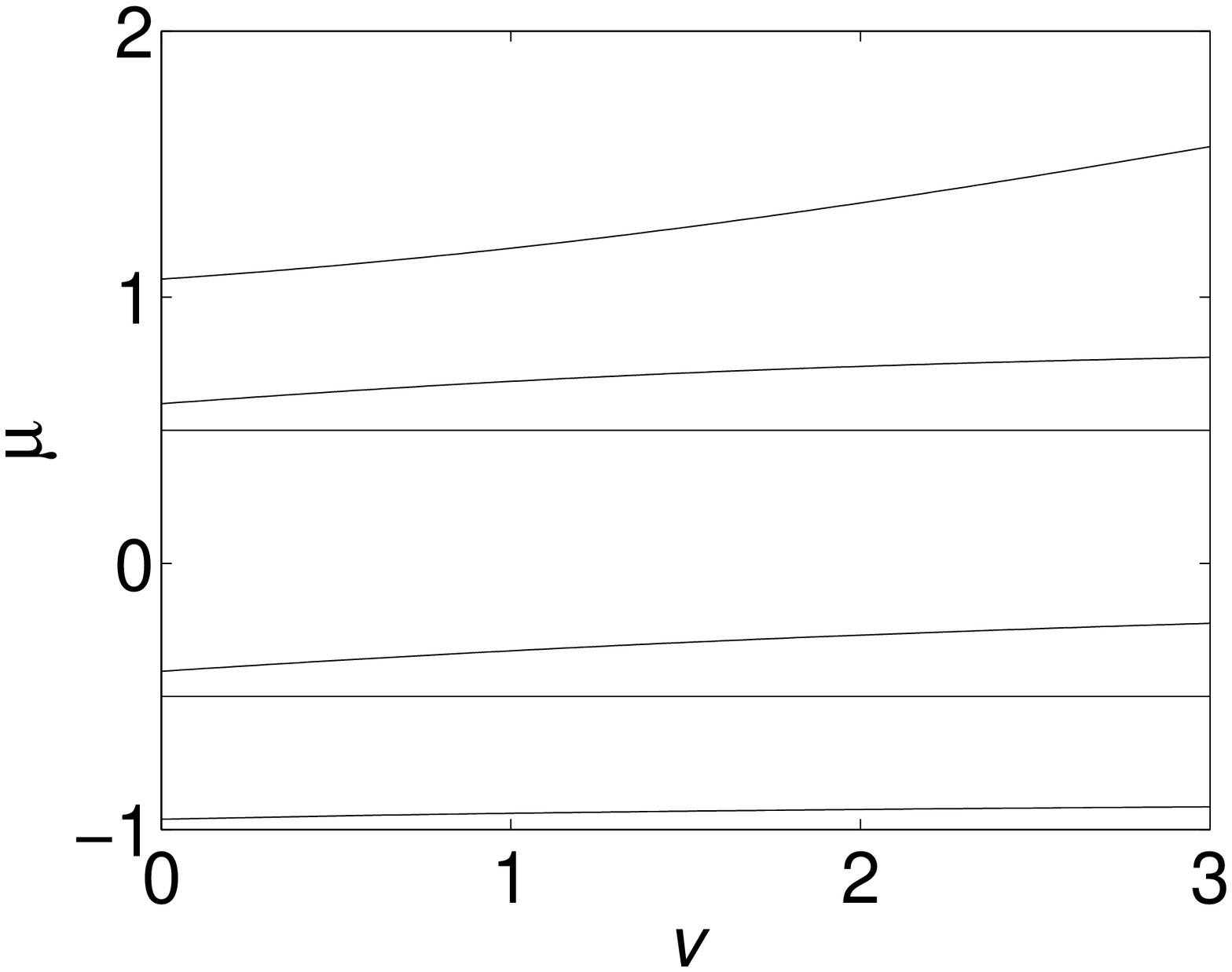}
\end{center}
\caption{ In both panels $U =\! 0.05$ and $M =\! N = \! 6$. 
Left panel: distribution of $\mu_k$ and $c_k = \cos (y_k)$ as functions of $\tau$ for
$V_0 =\! 0.4$. Right panel: distribution of $\mu_k$ and $c_k $
as functions of $v$ for $T = \! 0.5$. SP energies are given by $-2\mu_k T$
and $-2T c_k$.}
\label{fig1}
\end{figure}

The dependence of such energies from $\tau= T/UN$ and $v= V_0/UN$ is illustrated in figure \ref{fig1}.
The formation of energy doublets predicted by equation (\ref{sp1}) is well visible in the
left panel for large $\tau$: for each $k$, $-\lambda_k/(2T)$ and $- \cos (y_k)$
are separated by a small gap ${w}/{T M} <<1 $ if $w = V_0 +UN$ is small enough.
Furthermore, one easily recognizes the solution described by formula (\ref{sp2}) 
due to its diverging behavior $\mu \simeq \sqrt {1+ {w^2}/{4T^2} }$ for $\tau \to 0$ and $\mu \to 1$
for large $\tau$. In the right panel of figure \ref{fig1} the $\mu$ values forming doublets
at $v \simeq  0$ tend to the more uniform distribution described by equation (\ref{sp2})
as $v$ (and thus $w$) increases. 

In the non interacting limit $U=0$ Hamiltonian (\ref{HVo}) reduces to
$H  = C_N - V_0\, n_{0} - T \sum_i ( a^+_i a_{i+1} + a^+_{i+1} a_{i} )$ with $w = V_0$
and the doublet structure becomes the distinctive
feature of SP energies provided $w = V_0$ is small enough. This case is only apparently correct
in that the procedure whereby model (\ref{HVo}) (and $H_w$) is attained is not
justified: a weak $U$ does not support the boson localization at $j=0$. 
Then the case when $U$, $V_0$ (and thus $w = V_0 +UN$) are weak is not acceptable even if ${w}/{T M} <<1$ 
is still valid. 
%
%%%%%%%%%%%%%%%%%%%%%%%%%%%%%%%%%%%%%%%%%%%%%%%%%%%%%%%%%%%%%%%%%%%%%%%%%%%%%%%%%
%%%%%%%%%%%%%%%%%%%%%%%%%%%%%%%%%%%%%%%%%%%%%%%%%%%%%%%%%%%%%%%%%

The SI regime, where $U$ is large and $T/w << M/4$, 
features SP energies $-\lambda_k$ that are small deviations from $-2T \cos ( {\bar y}_k)$
``far" from $-2T \cos ( {y}_k)$. In this case no doublet structure is found
(see the left panel of figure \ref{fig1} for small $\tau$)
labels $y_k$, ${\bar y}_k$ of SP energies being uniformly distributed 
in interval $y\in [0, \pi]$. 
This case includes the situation where $U$ is weak or zero but well depth $V_0$ is
large enough to sustain the boson localization on which our approximation relies.
%
%

%%%%%%%%%%%%%%%%%%%%%%%%%%%%%%%%%%%%%%%%%%%%%%%%%%%%%%%%%%%%
\begin{figure}
\begin{center}
\includegraphics[width=6.0cm]{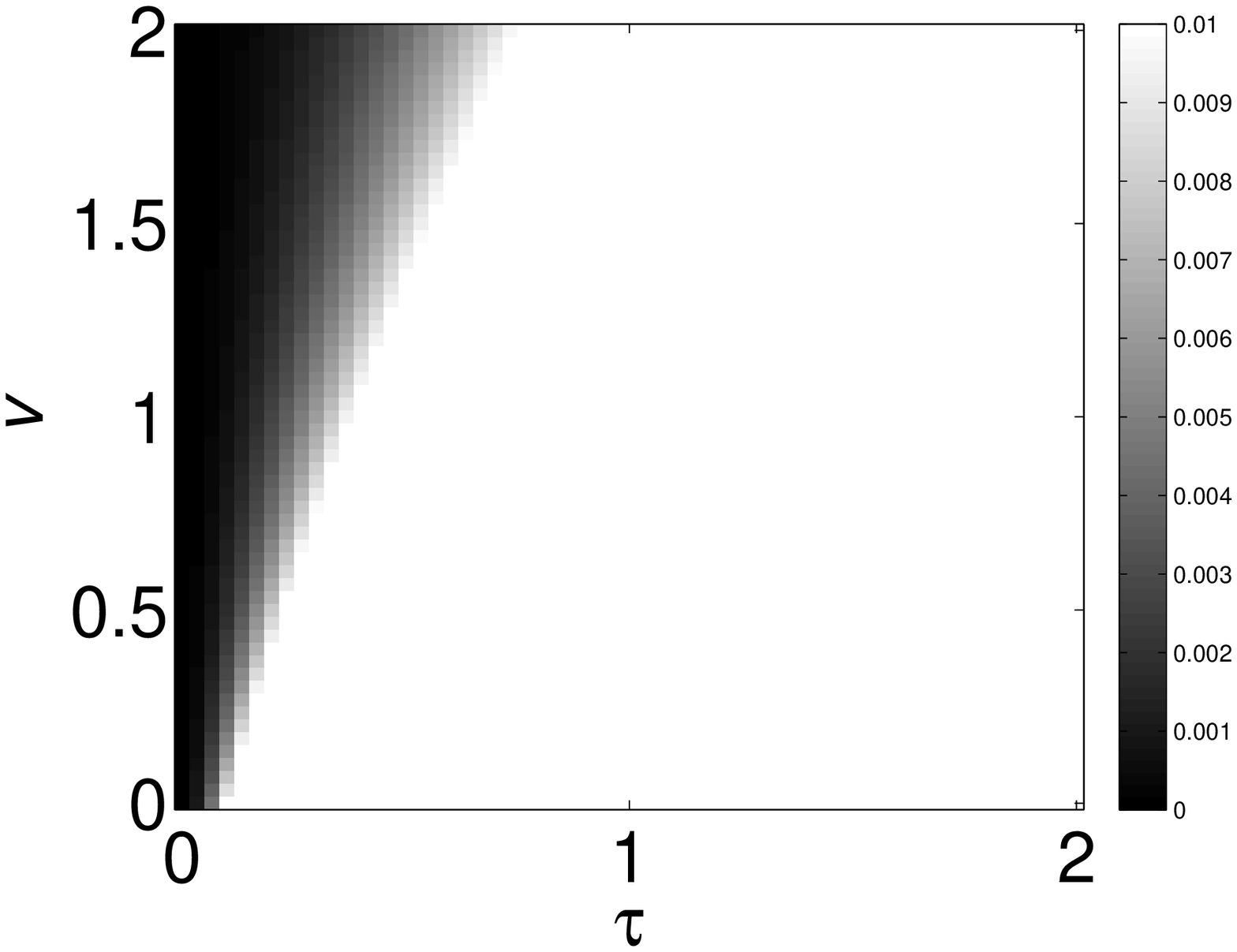}
\includegraphics[width=6.0cm]{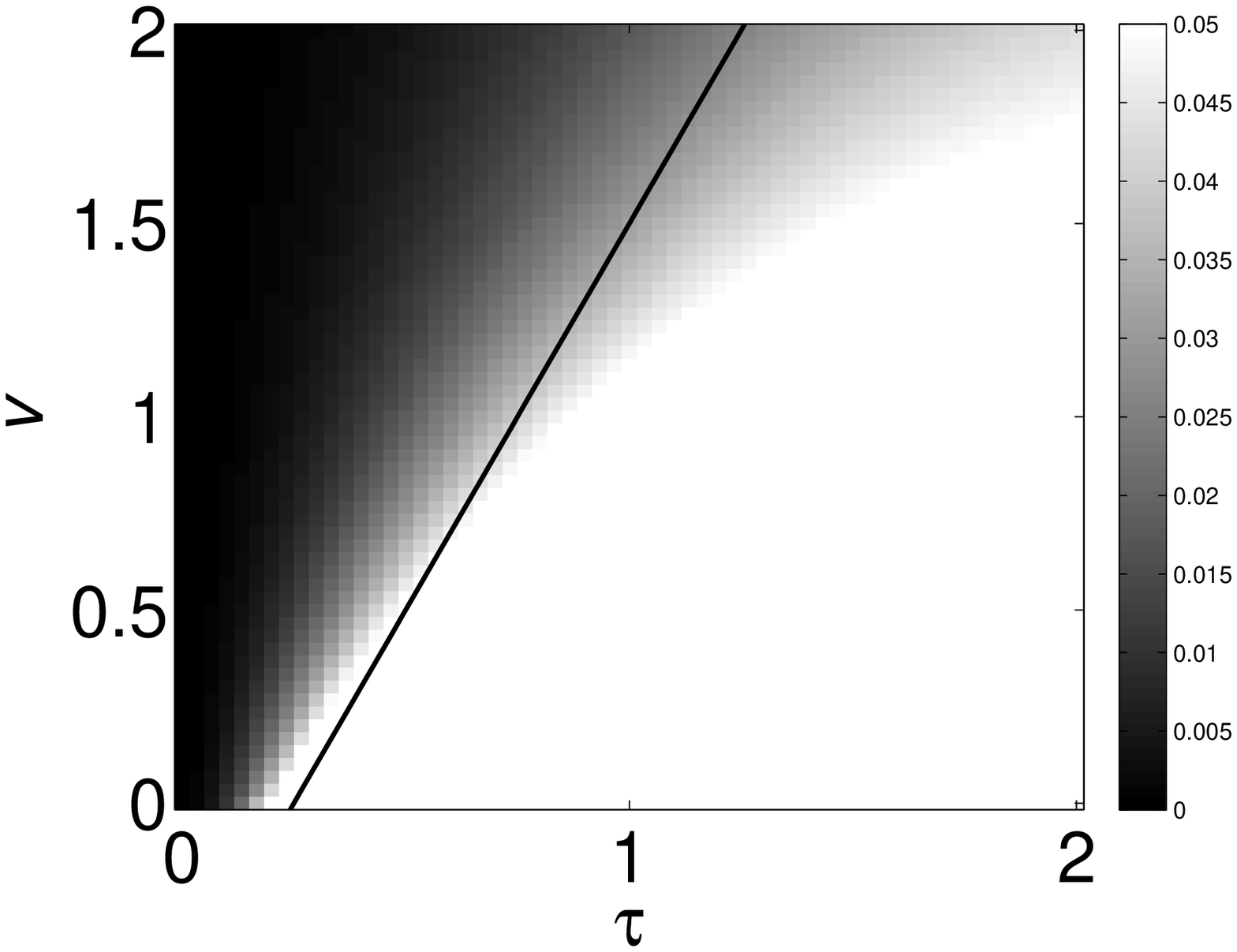}
\end{center}
\caption{
Relative error $|E_0 - E_{\mathrm{gs}}|/ E_0 $ given
in terms of the approximate ground-state energy $E_{\mathrm{gs}}$ and of its exact value $E_0$
for $M =\! N = \! 6$. The grey scale describes the relative error whose maximum is 1 
$\%$ (5 $\%$) in the left (right) panel.
}
\label{fig3}
\end{figure}
%%%%%%%%%%%%%%%%%%%%%%%%%%%%%%%%%%%%%%%%%%%%%%%%

\begin{figure}
\begin{center}
\includegraphics[width=7.0cm]{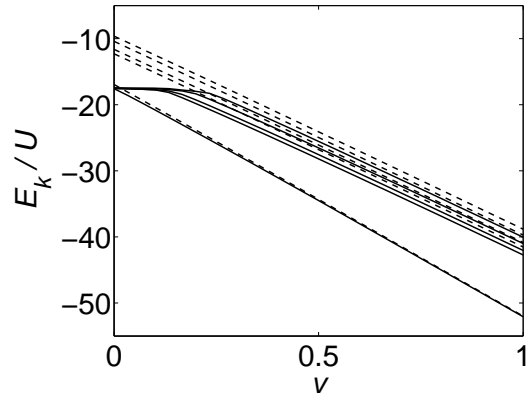}
\end{center}
\caption{
Dependence on $v= V_0/UN$ of the first five energy eigenvalues
for $\tau=1/6$, $M =\! N = \! 6$. Continuous lines describe exact eigenvalues obtained numerically,
while dashed lines are obtained from formula (\ref{ener}).}
\label{fig4}
\end{figure}

In figure \ref{fig3}, the ground-state energy $E_{\mathrm{gs}}$ obtained from equation (\ref{ener}) is compared
with the exact ground-state energy $E_0$, evaluated numerically, through the relative error $|E_{\mathrm{gs}}-E_0|/E_0$.
In both panels, extended regions of plane $(\tau, v )$ appear to involve an almost negligible relative error.
In the right panel, the straight line $v = 2\tau -1/2$ roughly separates the region
where the approximation of $E_0$ through $E_{\mathrm{gs}}$ is extremely good from the one where it becomes unsatisfactory.
Such a separatrix can be obtained with a simple semiclassical argument: assume that operators $a_i$ and $a_i^+$ in 
Hamiltonian (\ref{HV}) are replaced by complex variables $z_i$ and $z_i^*$. Its semiclassical counterpart thus
reads $H = -\sum_i U|z_i|^4/2 - V_0 |z_0|^2 - T \sum_i (z^*_i z_{i+1} + z_i z^*_{i+1} )$.
The latter, depending on the value of $\tau$ and $v$, exhibits two possible ground-state configurations: 
$z_0 = \sqrt N$, $z_i \simeq 0$ for $i\ne 0$ (solitonlike state) and $z_i \simeq \sqrt {N/M}$ (uniform state)
giving
$$
E'_0 = -N^2 U \left ( \frac{1}{2}  + v \right ), \,  
E''_0 = -N^2 U \left ( \frac{1}{2M}  +  \frac{v}{M} +2 \tau \right ) ,
$$
respectively. The situation where $E'_0 > E''_0$ and hence the ground state is uniform (SF regime)
entails the inequality $v < 2\tau -1/2$. Then the region where the approximation $E_0 \simeq E_{\mathrm{gs}}$
is no longer satisfactory essentially identifies with the SF regime. Figure \ref{fig4} describes the
first five energy eigenvalues in the range  $v\in [0,2]$ for $\tau=1/6$. Eigenvalues (\ref{ener})
well approximate qualitatively exact eigenvalues obtained numerically for $v> 0.3$,
consistent with figure \ref{fig3}. In particular, $E_0$ is almost
indistiguishable from $E_{\mathrm{gs}}$.
%
%%%%%%%%%%%%%%%%%%%%%%%%%%%%%%%%%%%%%%%%%%%%%%%%
%%%%%%%%%%%%%%%%%%%%%%%%%%%%%%%%%%%%%

\begin{figure}
\begin{center}
\includegraphics[width=6.0cm]{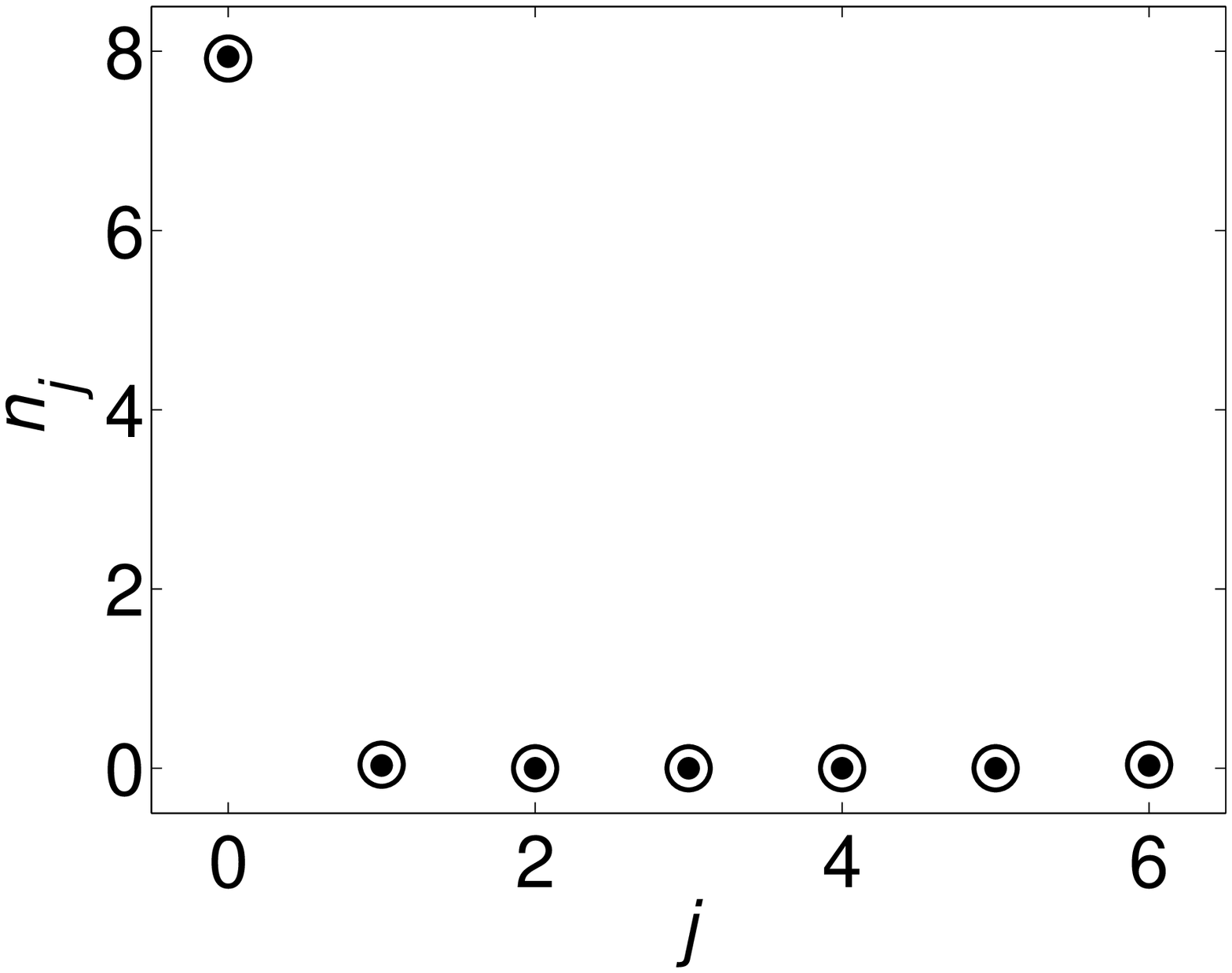}
\includegraphics[width=6.0cm]{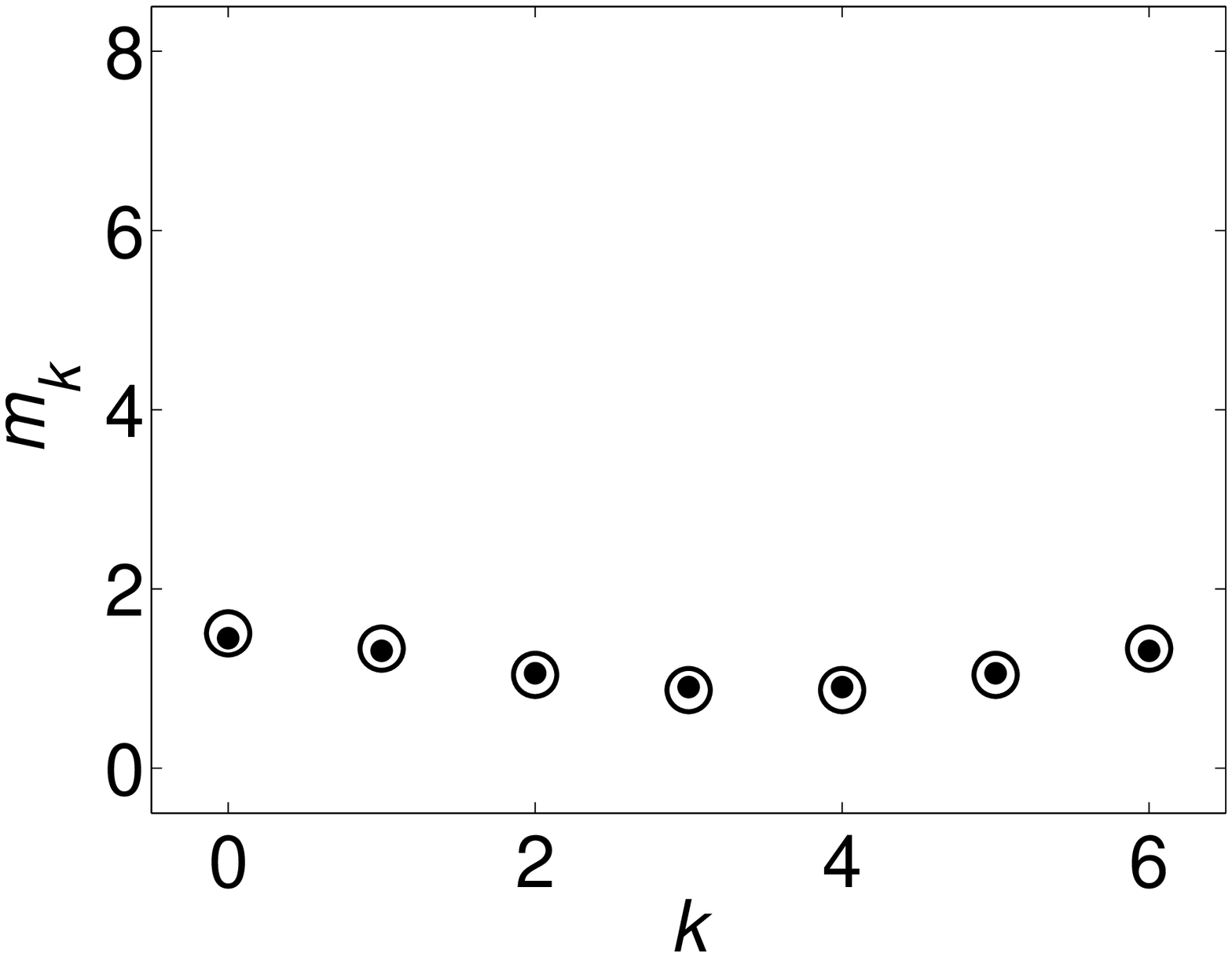}
\end{center}
\caption{
Space distribution $n_j= \langle a^+_j a_j \rangle$ (left panel) and momentum distribution
$m_k = \langle b^+_k b_k \rangle$ (right panel) of bosons in the ground state for 
$T =\! 0.5$, $V_0 =\! 0.1$, $U =\! 1$, $M =\! 7$, $N = \! 8$. 
Symbols $\bigcirc$ and $\bullet$ describe the mode occupation relevant to the exact ground state
and to its approximated form (\ref{GS}), respectively, showing an excellent agreement.
}
\label{fig2}
\end{figure}

To test our approximation scheme we further calculate 
the boson distribution in the ambient space corresponding to the ground state
by exploiting the properties of SU($M$) coherent states \cite{BP}.
% The boson distribution in the ambient space corresponding to the ground state
% can be easily calculated by exploiting the properties of SU($M$) coherent state \cite{BP}.
To this end we reformulate operator $D^+_0$ in $|\mathrm{GS} \rangle$ in terms
of space modes. This procedure, developed in appendix \ref{A4}, gives
$$
n_j =
\langle \mathrm{GS}| a^+_j a_j |\mathrm{GS} \rangle = N |\xi_j|^2
= 
N\, \frac{ 2\, {\rm ch}^2 [(M/2-j)y] }{ M + {\rm sh} (M y) \, {\rm coth} y  }
$$
obeying the normalization condition  $\sum_i |\xi_j|^2=1$, together with the boson distribution
$m_k = \langle \mathrm{GS}| b^+_k b_k |\mathrm{GS} \rangle $ $= N |x_k|^2$ among momentum modes
%(see appendix \ref{A4})
where
$$
|x_k|^2
=
\frac{ 2 \, {\rm sh}^2 (M y/2) \, {\rm sh}^2 y }{ M + {\rm sh} (M y) \, {\rm coth} y  }
\, \frac{ 1 }{ ({\rm ch} y - c_k)^2} \, .
$$ 
In these two equations $e^y = (w/2T + \sqrt{ 1+ w^2/4T^2 })/2 $ 
if $T/w<<M/4$, as stated by the second equation in formula (\ref{Ssing}).
In the SI regime, where $UN >> T$, one has $e^y = w/2T >>1$
so that $n_j = N |\xi_j|^2 \simeq N \exp ( \, |2j-M|y -My)$. 
%
%$e^y \simeq w/2T >>1$ 
%
Note that index $j$ ranges in $[0, M-1]$ since $j=M$ is equivalent to $j=0$ in the
ring geometry. Then the maximum occupation in the lattice is reached at
$j=0$, where the boson-population peak is expected, while the minimum occupation
is found at $j=M/2$. Figure \ref{fig2} shows how $n_j $ is in an excellent agreement
with the boson space distribution supplied by the exact (numeric) calculation of the ground state
in regime $T/w<<M/4$. Figure \ref{fig2} (right panel) 
confirms as well the validity of momentum-mode distribution $\langle m_k \rangle$ in the same regime.

%%%%%%%%%%%%%%%%%%%%%%%%%%%%%%%%%%%%%%%%%%%%%%%%%%%%%%%%%%%%%%%%%%%%%
%
%%%%%%%%%%%%%%%%%%%%%%%%%%%%%%%%%%%%%%%%%%%%%%%%%%%%%%%%%%%%%%%%%%%%%%%%%%%%%%%%%%%%%%%%%%%%%%%%
%%%%%%%%%%%%%%%%%%%%%%%%%%%%%%%%%%%%%%%%%%%%%%%%%%%%%%%%%%%%%%%%%%%%%%%%%%%%%%%%%%%%%%%%%%%%%%%%
%%%%%%%%%%%%%%%%%%%%%%%%%%%%%%%%%%%%%%%%%%%%%%%%%%%%%%%%%%%%%%%%%%%%%%%%%%%%%%%%%%%%%%%%%%%%%%%%
%%%%%%%%%%%%%%%%%%%%%%%%%%%%%%%%%%%%%%%%%%%%%%%%%%%%%%%%%%%%%%%%%%%%%%%%%%%%%%%%%%%%%%%%%%%%%%%%
%
\section{Spectrum of the SF regime}
\label{sfreg}

To determine the spectrum of model (\ref{HV}) within the SF regime we implement the standard
Bogoliubov scheme involving the formulation of $H$ within
momentum-mode picture. The full diagonal form of $H$ is achieved
by means of a three-step procedure the first step of which consists in replacing $a_j$ with
$a_j = \sum_k  b_k\, {e^{i{\tilde k}j } }/{\sqrt M}$
and $b^+_0b_0$ with $b^+_0b_0 = N - \sum_{k\ne 0} b^+_k b_k$.
Hamiltonian (\ref{HV}) becomes 
\begin{equation}
H \simeq   H_2 - V_0 \, \sqrt {n} ( B^+ + B ) -\Lambda
\label{HbogoV}
\end{equation}
where $n = N/M$, the quadratic part of $H$ reads
$$
H_2 \simeq  \sum_{k \ne 0} \Bigl [ g_k b^+_kb_k 
- \frac{Un}{2}  \, \left ( b^+_{-k} b^+_k  +  b_{k}  b_{-k} \right ) \Bigr ] -V_0 B^+ B 
$$
with $g_{k} = {V_0}/{M} + e_k -Un$, $e_k =2T [1-\cos({\tilde k})]$, and
$$
\Lambda = \frac{U}{2M}  N(N-1)\, +\, 2TN  + nV_0 \, ,\quad B =\sum_{k\ne 0} \, \frac{b_k}{\sqrt M}\, .
$$ 
The presence of the quadratic term $H_2$ in $H$ suggests 
that linear term $V_0 \sqrt {n} ( B^+ + B )$ can be eliminated through the combined action of 
$k$-dependent displacement operators
$$
T_k = e^{z_k b^+_k - z^*_k b_k} \, , \quad T_k b_k T^+_k = b_k - z_k\, .
$$
As shown in appendix \ref{A5}, after implementing the unitary transformation $H \to {\cal H} = R^+ H R $
with $R = \prod_{k\ne 0} T_k$, choice (\ref{xk}) of undetermined parameters $\eta_k$ provides
$$
{\cal H} = H_2 - C \quad ({\rm see \, formula} \, (\ref{qH}))
$$
with $C= \Lambda +\Phi$ and $\Phi = n V_0 /(1+S)$. The nice property of $\cal H$ is that
it can be taken into a diagonal form by means of relatively simple calculations. 
Appendix \ref{A5} illustrates the second step of our procedure which consists in separating
$H_R$ in two independent parts by exploiting again operators
$f_k = (b_k -b_{-k})/{\sqrt 2}$, $F_k = (b_k +b_{-k})/{\sqrt 2}$ given in formula (\ref{newop}). 
One finds ${\cal H} = {\cal H}_{f} + {\cal H}_F$ where
$$
{\cal H}_f = -C  
+\! \sum^S_{k=1} \Bigl [
g_{k}  f^+_{k} f_k + \frac{Un}{2} \Bigl ((f^+_{k} )^2 + f_{k}^2  \Bigr )  \Bigr ] \, ,
$$
$$
{\cal H}_F = 
\! \sum^K_{k =1} \Bigl [ g_{k}  F^+_{k} F_k 
-\frac{Un}{2} \Bigl ( (F^+_{k} )^2  + F_{k}^2 \Bigr ) \Bigr ] -V_0 B^+ B \, ,
$$
where $B = \sum^K_{k = 1} r_k F_k/{\sqrt M}$, $r_k$ is defined after equation (\ref{newop}),
and parameters $S$ and $K$ have been defined in formulas (\ref{Sran}) and (\ref{Kran}). 
Hamiltonian ${\cal H}_f$ is easily diagonalized through the procedure described in \cite{Sol1}, 
\cite{Sol2}. Since 
$$
J^z_k = \frac{2f^+_{k} f_k +1}{4} ,\,\, J^x_k = \frac{(f^+_{k} )^2 + f^2_{k} }{4}
, \,\, J^y_k = \frac{(f^+_{k} )^2 - f^2_{k} }{4i} ,
$$
are, for each $k$, the
generators of an algebra su(1,1) obeying commutators
$[J^x_k, J^y_k ] = -2iJ^z_k$, $[J^y_k, J^z_k ] = 2iJ^x_k$ and $[J^z_k, J^x_k ] = 2iJ^y_k$, 
then the unitary transformation 
$$
D_k J^z_k D^+_k = J^z_k {\rm ch} \alpha_k  + J^x_k {\rm sh} \alpha_k
\, ,\quad
D_k = e^{-i \alpha_k J^y_k}\,  ,
$$ 
allows one to diagonalize ${\cal H}_f$. We then rewrite ${\cal H}_f$ as 
$$
{\cal H}_f = {\sum}^S_{k=1} \Bigl [ \, g_{k}  (2J^z_{k}-1/2) + 2Un J^x_{k} \Bigr ] -C \, .
$$
whose diagonal form ${\rm H}_f$ is achieved by means of transformation $D = \Pi_k D_k $
$$
D {\rm H}_{f} D^+ =
D \Bigl [ \,  {\sum}^S_{k=1} ( 2\nu_{k}  J^z_{k} - {g_{k}}/{2} ) -C \Bigr ] D^+ = {\cal H}_f \, ,
$$
if conditions $g_k = \nu_k {\rm ch} \alpha_k$ and $Un = \nu_k {\rm sh} \alpha_k$ are imposed.
In ${\rm H}_{f} $ parameters $\nu_k$ read
\begin{equation}
\! \nu_k  =  \! \sqrt { g_k^2 - U^2n^2 } 
\! = \! \sqrt { \! \Bigl ( \! {V_0}/{M} +e_k \!- \! Un \Bigr )^2 - U^2n^2 }
\label{nu}
\end{equation}
giving the energy eigenvalues relevant to ${\cal H}_f$
$$
E_f ({\vec p} \,)
= {\sum}^S_{k=1} \left [ \,\nu_{k} (p_k +1/2) -{g_{k}}/{2}  \right ] -C 
$$
in the Fock-space basis formed by states $| {\vec p}\,\rangle = \prod^S_{k=1} | p_k \rangle$
where $f^+_k f_k| p_k \rangle= p_k| p_k \rangle$, $p_k = 0,1,2 ...$ .
The third and last step of the diagonalization process (see appendix \ref{A5}) 
concerns ${\cal H}_F$ which can be rewritten as
$$
{\cal H}_F = \! {\sum}^K_{k,h =1} G_{kh} F^+_{k} F_h 
-\frac{Un}{2}  {\sum}^K_{k =1} \Bigl ( F_{k}^2 + \mathrm{H.C.} \Bigr ) \, ,
$$
with $G_{kh} =  g_{k}\delta_{kh} -V_0 {r_h r_k}/{M}$.
One can implement the same scheme applied for diagonalizing the component $H_F$ of $H_w$
in the SI regime. To this end we introduce new operators $C_\ell = \sum_h f_{h \ell} F_h$
and $C^+_\ell = \sum_h f_{h \ell} F^+_h$ such that $[C_\ell , C^+_m] = \delta_{\ell m}$.
Parameters $f_{h \ell}$ are undefined elements of an orthogonal matrix which can be exploited to
take $G_{kh}$ into a diagonal form. Appendix \ref{A5} illustrates the calculations whereby
the $C_k$-dependent final form 
$$
{\cal H}_C
= {\sum}^K_{\ell =1} \theta_{\ell} C^+_{\ell} C_\ell
-\frac{Un}{2} {\sum}^K_{k =1} \Bigl ( (C^+_{k})^2 + C_{k}^2 \Bigr )
$$
of Hamiltonian ${\cal H}_C$ is found
together with $f_{h \ell}$ and $\theta_\ell$. The latter (see equations (\ref{aresul1})) are given by
\begin{equation}
f_{h \ell}  = - \frac{V_0}{M}  \frac{r_h Y_\ell}{ \theta_\ell -g_h }
\, , \,\,
1 = - \frac{V_0}{M} \sum^K_{h=1}  \frac{r^2_h}{\theta_\ell -g_h } \, ,
\label{autovtheta}
\end{equation}
where $Y_\ell = \sum^K_{k=1} r_k f_{k \ell} $ is determined in appendix \ref{A5}.
Fortunately, Hamiltonian ${\cal H}_C$ exhibits the same algebraic structure 
of ${\cal H}_f$. Then, also in this case, its diagonal form ${\rm H}_C$ 
is connected to ${\cal H}_C$ by 
$$
W {\rm H}_C W^+ =
W \Bigl [ \, 
{\sum}^K_{h=1} \left ( 2 \eta_{h}  V^z_{h} -{\theta_{h}}/{2} \right )  \Bigr ] W^+ = {\cal H}_C \, ,
$$
in which $W = \Pi_k W_k$ is a unitary transformation whose factors are defined as
$W_k = \exp (i \beta_k V^y_k )$.
Similar to $J^z_k$, $J^x_k$ and $J^y_k$ operators $V^z_k$, $V^x_k$ and $V^y_k$ are,
for each $k$, generators of an algebra su(1,1) written in terms of $C_k$ and $C^+_k$ instead of $f_k$,
$f^+_k$. In particular, $V^z_{k} = (2C^+_k C_k +1)/4$. 
Conditions $\theta_h = \eta_h {\rm ch} \beta_h$ and $Un = \eta_h {\rm sh} \beta_h$ ensure that
$W {\rm H}_C W^+ ={\cal H}_C$ and provide the definitions
\begin{equation}
\eta_k = \sqrt { \theta_k^2 - U^2n^2 } 
\, ,\quad 
{\rm th} \beta_k = {Un}/{\theta_k }\, .
\label{eta}
\end{equation}
Thanks to parameters $\eta_h$, we easily identify
the energy eigenvalues relevant to ${\cal H}_C$ (and thus to ${\cal H}_F$)
$$
E_F ({\vec q} \,)
= {\sum}^K_{k=1} \left [  \, \eta_{k} ( q_k +1/2 ) -{\theta_{k}}/{2}  \right ] 
\, ,\quad q_k = 0,1,2 ...
$$
in the Fock-space basis formed by states $| {\vec q}\,\rangle = \Pi_k | q_k \rangle$
where $C^+_k C_k| q_k \rangle= q_k| q_k \rangle$.
Summarizing, the eigenvalues of total Hamiltonian ${\rm H}_f + {\rm H}_C$ are
\begin{equation}
E ({\vec p}, {\vec q} \,) = E_F ({\vec q} \,) + E_f ({\vec p} \,)\, ,
\label{enerSF}
\end{equation}
corresponding to eigenvectors $| {\vec p}, {\vec q} \rangle =  \Pi^S_{k=1} | p_k \rangle  \Pi^K_{h =1} | q_h \rangle$. 
Energies (\ref{enerSF}) provide as well the spectrum of Hamiltonian 
(\ref{HbogoV}), 
%unitarily equivalent to ${\rm H}_f + {\rm H}_C$, 
whose eigenvalue problem is
\begin{equation}
H |E ({\vec p}, {\vec q} \,)\rangle = E ({\vec p}, {\vec q} ) |E ({\vec p}, {\vec q} \,)\rangle
\label{eigenvec}
\end{equation}
where $|E ({\vec p}, {\vec q} \,)\rangle = RDW \,| {\vec p}, {\vec q} \rangle$.
This concludes the diagonalization
process whose validity is supported by the fact that one has
$\{ \theta_\ell \} \equiv \{ g_k \}$, $f_{k\ell} \equiv \delta_{k\ell}$, $C_k =  F_k$
and $x_k =0$, for $V_0 \to 0$. In this case the usual scenario relevant to the Bogoliubov scheme 
where $\eta_k \equiv \nu_k$ is recovered and no splitting effect, causing $\eta_k \ne \nu_k$, is observed. 
%

%%%%%%%%%%%%%%%%%%%%%%%%%%%%%%%%%%%%%%%%%%%%%%%%%%%%%%%%%%%%%%%%%%%%%%%%%%%%%%%%%%%%%%%
\subsection{Discussion}
\label{qp}

An important aspect of the diagonalization scheme leading to quasi-particle energies
(\ref{nu}) and (\ref{eta}) concerns the range of parameters in which it should be valid. This
is related to the conditions ensuring that quasi-particle energies are real and
positive. The first condition is
\begin{equation}
g_k - Un = V_0/M + e_k -2Un > 0 \, ,
\label{posit1}
\end{equation}
which, being $g_k + Un = V_0/M + e_k > 0$ for any $k$, implies that 
$g_k^2 - U^2n^2 > 0$ in equation (\ref{nu}). The second one is
\begin{equation}
\theta_k + Un > 0   \, ,\quad \theta_k - Un > 0,
\label{posit2}
\end{equation}
ensuring that $\theta_k^2 - U^2n^2 $ is positive in equation (\ref{eta}). By using parameters
$\tau$ and $v$, inequality (\ref{posit1}) reduces to
\begin{equation}
{\tau } > {(1 -v/2)} [\, 2M \sin^2 (\pi k/M)]^{-1}
\label{posit3}
\end{equation}
which essentially reproduces the well-known condition establishing the parameter-$\tau$ 
interval in which the Bogoliubov approximation is valid. In the worst case ($k=1$) this
inequality gives ${\tau}> (1 -v/2) {M}/{(2 \pi^2)}$ for large $M$. The novelty here is represented by 
factor $1-v/2$ showing that such an interval is enlarged because $v \ne 0$ due to the presence of
local potential $V_0$.

Inequalities (\ref{posit2}) involve a more complicated situation. Parameters $\theta_\ell$ are solutions 
of equation (\ref{autovtheta}) which, being 
$g_k = V_0/M +2T (1-c_k)- Un$ can be rewritten as 
\begin{equation}
\frac{2T M}{V_0} = \!\sum_{k\ne 0}  \!
\left [ \Bigl ( 1 - \frac{UN-V_0}{2TM}  - c_k \Bigr ) - \mu \right ]^{-1}\! = {\cal F} (\mu)
%\frac{1}{\mu - \left ( 1 - \frac{UN-V_0}{2TM}  - c_k \right )}  = {\cal F} (\mu)
\label{automu}
\end{equation}
with $\mu \equiv {\theta }/{2T }$ (we drop index $\ell$ of $\theta_\ell$ which is viewed as
a continuous variable). In general, such an equation can be solved only numerically and
no simple condition such as inequality (\ref{posit3}) is available in this case.

%%%%%%%%%%%%%%%%%%%%%%%%%%%%%%%%%%%%%%%%%%%%%%%%%%%%%%%
\begin{figure}
\begin{center}
\includegraphics[width=7.5cm]{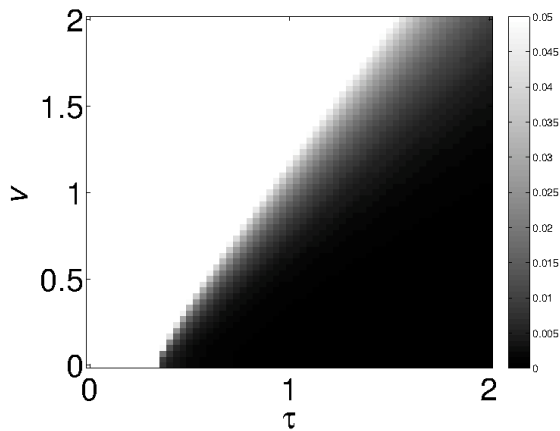}
\end{center}
\caption{
Relative error $|(E_0-E_{\mathrm{gs}})\,/\,E_0|$ given in terms of the
approximate ground-state energy $E_{\mathrm{gs}}$ and of its exact value
$E_0$ for $M=N=6$. The grey scale describes the relative error
whose maximum is $5\%$.
}
\label{fig_vtauSF}
\end{figure}
%%%%%%%%%%%%%%%%%%%%%%%%%%%%%%%%%%%%%%%%%%%%%%%%%%%%%%%%%%%%%%%%%%

The only exception is the regime ${2T M}/{V_0} >> 1$ (when, for example, $V_0$
is perturbative and/or $M$ is large enough)
in which approximate solutions can be found through an analytic approach (see Appendix \ref{A6}).
In this regime the number of solutions is expected to coincide
with the number $K$ of the asymptotes characterizing ${\cal F}( \mu) $ so that the quasi-particle
energy spectrum exhibits an evident doublet structure being $\eta_k \simeq \nu_k$.

After setting $\mu = V_0/(2TM) - Un/(2T) + 1 -\cos y $,
approximate solutions of $2TM/V_0={\cal F} (\mu)$ are found 
by substituting $y = y_k + \xi_k$ in its trigonometric version (\ref{trigoSF}).
The Taylor expansion of the latter to the second order in $\xi_k$
yields equation (\ref{AtrigoSF}) whose solutions (\ref{root1}), at fixed $M$ and with
$t= 8T/(MV_0)$ sufficiently large, reduce to $\xi_k = - {8}/{( t M^2 s_k )}$ entailing
$$
\theta_k = 2T \mu_k =  \frac{V_0}{M}-Un +2T (1-\cos(y_k+\xi_k))
\simeq g_k - \frac{2V_0}{ M } .
$$
These results show that the two conditions (\ref{posit2}), now expressed as
$g_k - 2V_0/M > \pm Un$ and therefore as
$$
4 \tau M \sin^2(\pi k/M) > v
\, , \quad
4 \tau M \sin^2(\pi k/M) -2 > v ,
$$ 
can be fulfilled for large enough $\tau$ and sufficiently small $v$ in plane $v$-$\tau$ 
even in the most restrictive case $k=1$. Of course
these inequalities supply a limited information on the range of validity of our scheme 
since they have been obtained in the limiting case $t >>1$. A complete information 
is provided by inequalities (\ref{posit2}) only through a systematic numerical study of
solutions $\theta_\ell$ when the model parameters are varied. This analysis is beyond the
scope of this work.

%%%%%%%%%%%%%%%%%%%%%%%%%%%%%%%%%%%%%%%%%%%%%%%%
\begin{figure}
\begin{center}
\includegraphics[width=6.0cm]{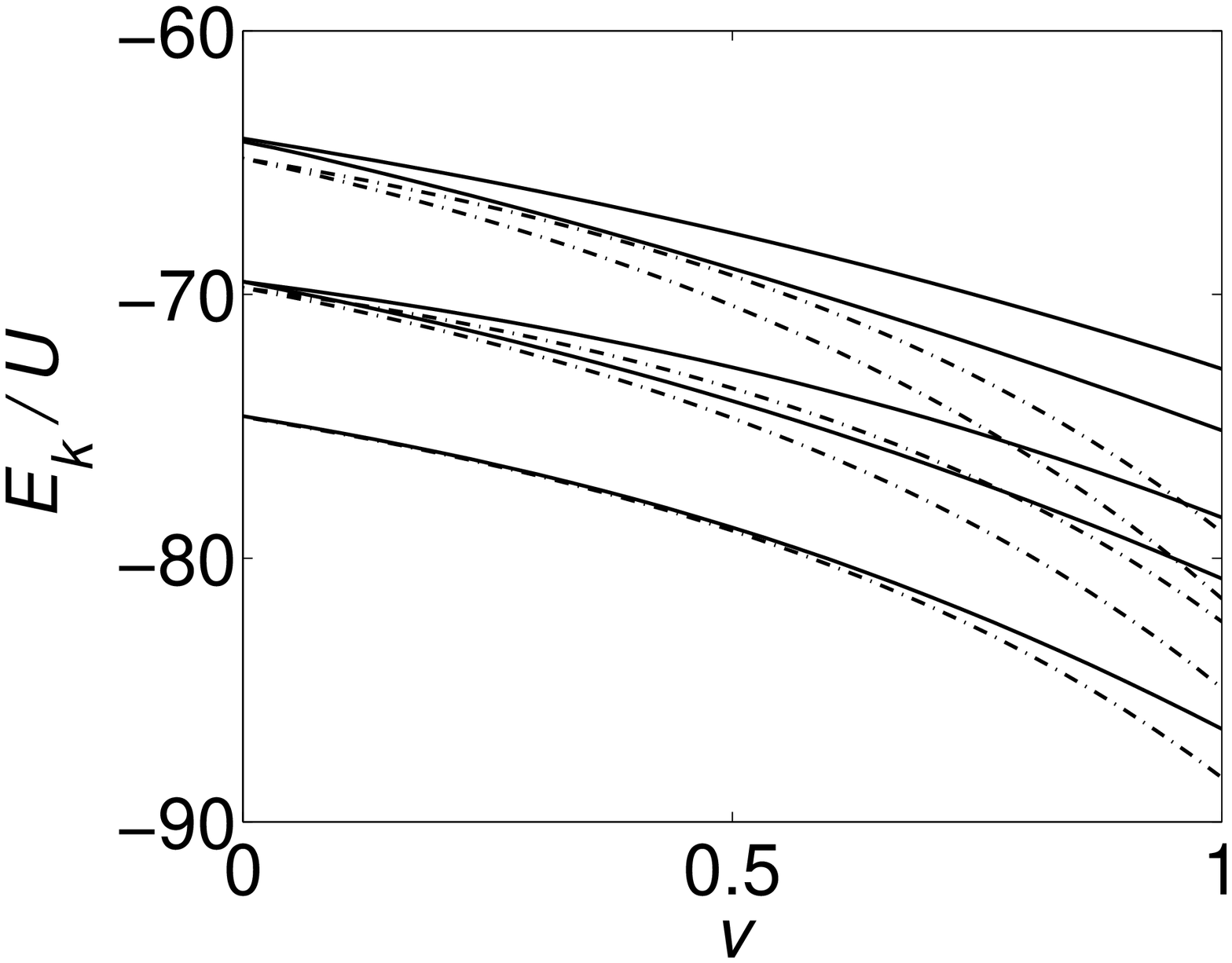}
\includegraphics[width=6.0cm]{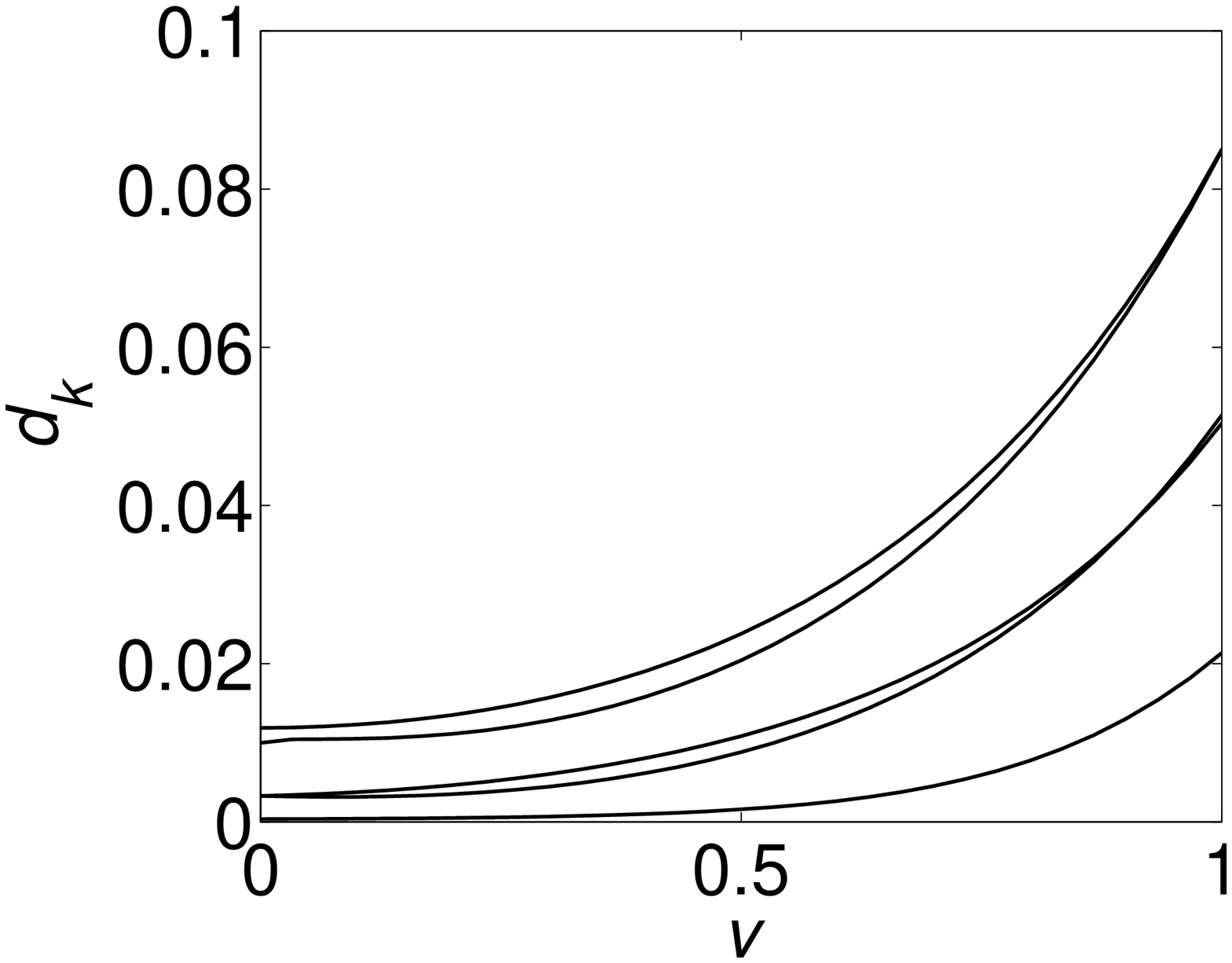}
\end{center}
\caption{Left panel: dependence on $v$ of the first five energy eigenvalues
for $\tau=1$, $M =\! N = \! 6$. Continuous lines describe exact eigenvalues denoted by
$E_k$ ($0 \le k \le 4 $) obtained numerically, while dashed lines describe eigenvalues 
$E^{\mathrm{ap}}_k$ obtained from formula (\ref{enerSF}). Right panel: $d_k  =|(E_k -E^{\mathrm{ap}}_k)/E_k |$
}
\label{fig5}
\end{figure}
%%%%%%%%%%%%%%%%%%%%%%%%%%%%%%%%%%%%%%%%%%%%%%%%%%%%%%%%%%

%%%%%%%%%%%%%%%%%%%%%%%%%%%%%%%%%%%%%%%%%%%%%%%%
\begin{figure}
\begin{center}
\includegraphics[width=6.3cm]{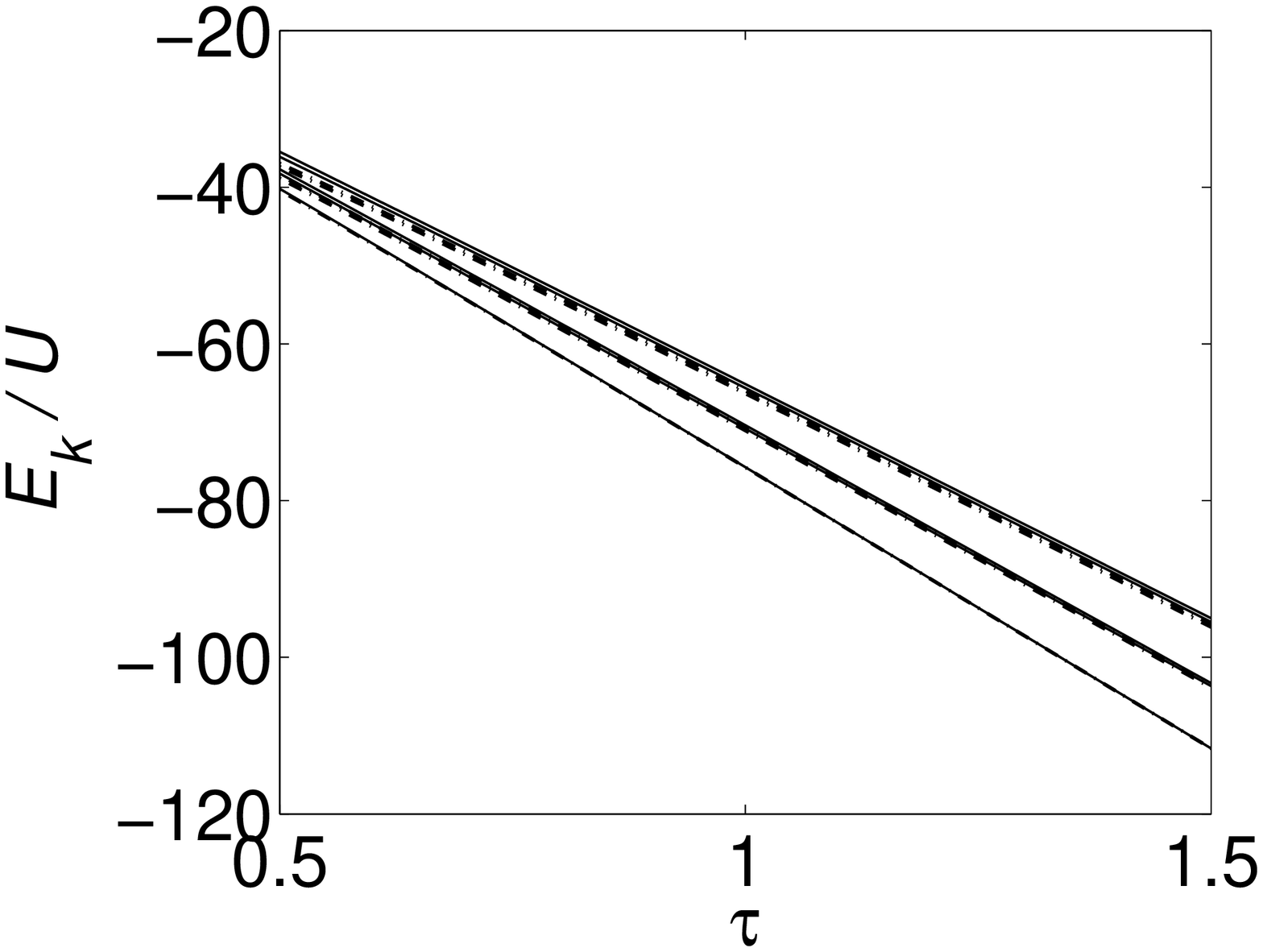}
\includegraphics[width=6.3cm]{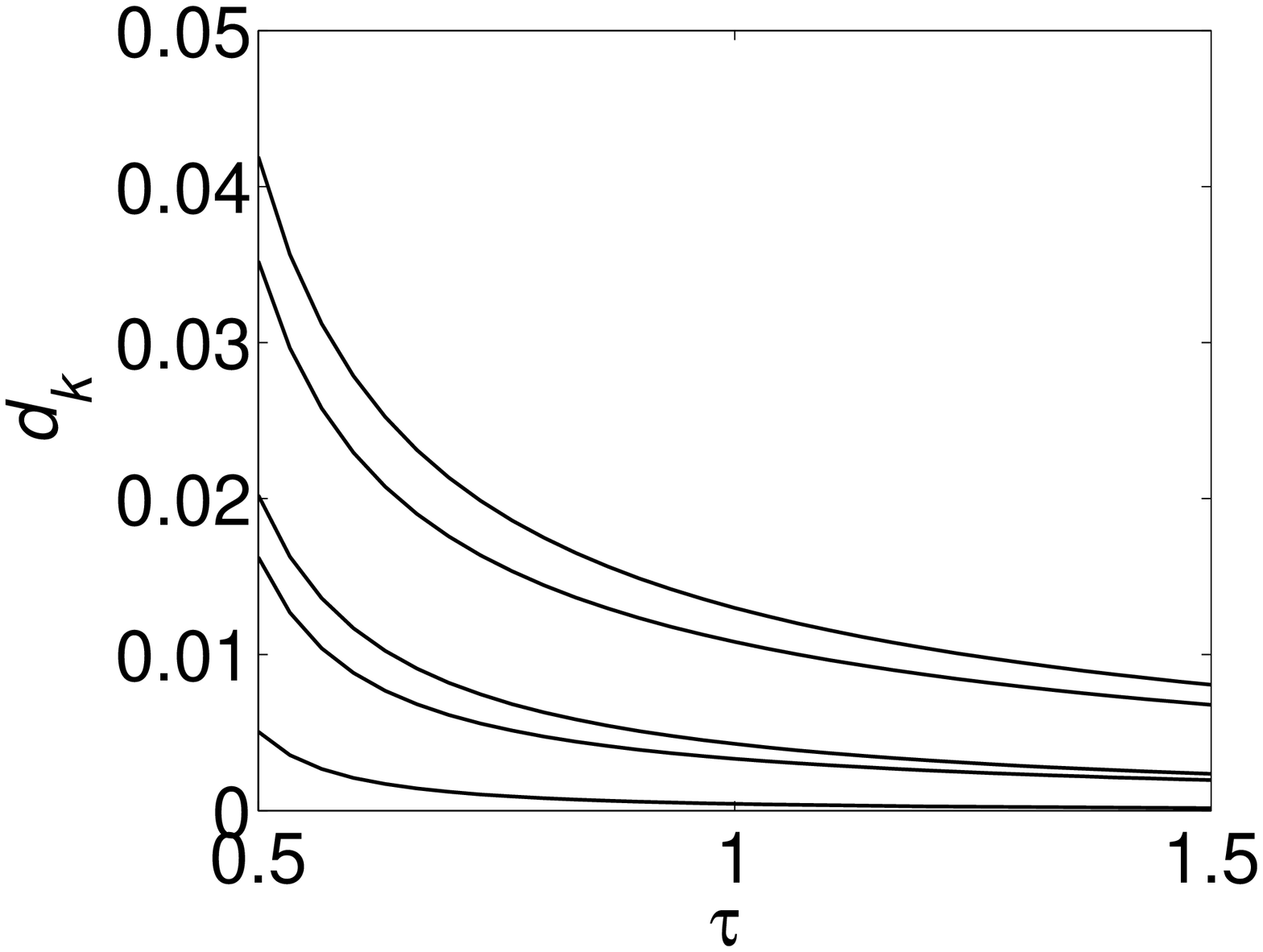}
\end{center}
\caption{Left panel: dependence on $\tau$ of the first five energy eigenvalues
for $v=1/6$, $M =\! N = \! 6$. Continuous lines describing $E_k$ and dashed lines 
describing $E^{\mathrm{ap}}_k$ have the same meaning as in figure \ref{fig5}. 
Right panel: $d_k  =|(E_k -E^{\mathrm{ap}}_k)/E_k |$
}
\label{fig6}
\end{figure}
%%%%%%%%%%%%%%%%%%%%%%%%%%%%%%%%%%%%%%%%%%%%%%%%%%%%%%%%
%
Here we limit our attention to some specific cases. In Figure \ref{fig_vtauSF}, the ground-state energy 
$E_{\mathrm{gs}}$, given by equation (\ref{enerSF}) with $p_k\,=\,q_h\,=\,0$ for all $k$ 
and $h$, is compared with the exact ground-state
energy $E_0$, evaluated numerically: in the figure the relative error $|(E_{\mathrm{gs}}\,-\,E_0)\,/\,E_0|$ 
is plotted as a function of $v$ and $\tau$. Also in this case, extended regions of the 
$(\tau;\,v)$ plane show an almost negligible relative error. In particular, at increasing values 
of $\tau$ one can see that the results provided by the Bogoliubov approximation become 
more precise, while the rise in $v$ implies a fast decrease in the effectiveness of the approximation
in the prediction of the ground-state energy. 

The scenario outlined in the previous paragraph is further confirmed by the numerical calculations 
of the first low-energy excited levels, illustrated in figure \ref{fig5} and \ref{fig6}.
In figure \ref{fig5}, where $\tau = 1$ while $v$ varies, the first five energy eigenvalues
$E_k$, $k\in [0,4]$ ($E_0$ corresponds to the ground state) obtained numerically can be compared 
with energies $E^{\mathrm{ap}}_k$ given by formula (\ref{enerSF}) within the Bogoliubov approximation. 
The agreement is, in general, extremely good for $v < 0.5$, and becomes excellent in the case of
the ground state, as shown by the relative error $d_k$ in the right panel of figure \ref{fig5}.
Figure \ref{fig6}, where $v=1/6$ while $\tau$ varies, shows that the agreement between
$E^{\mathrm{ap}}_k$ and $E_k$ is excellent for any value of $\tau > 0.5$. For $\tau < 0.5$ (this case is not shown)
the deviation of $E^{\mathrm{ap}}_k$ from $E_k$ becomes significant.
%
%%%%%%%%%%%%%%%%%%%%%%%%%%%%%%%%%%%%%%%%%%%%%%%%%%%%%%%%%%%%%%%%%%%%%%%%%%%%%%
\begin{figure}
\begin{center}
\includegraphics[width=6.0cm]{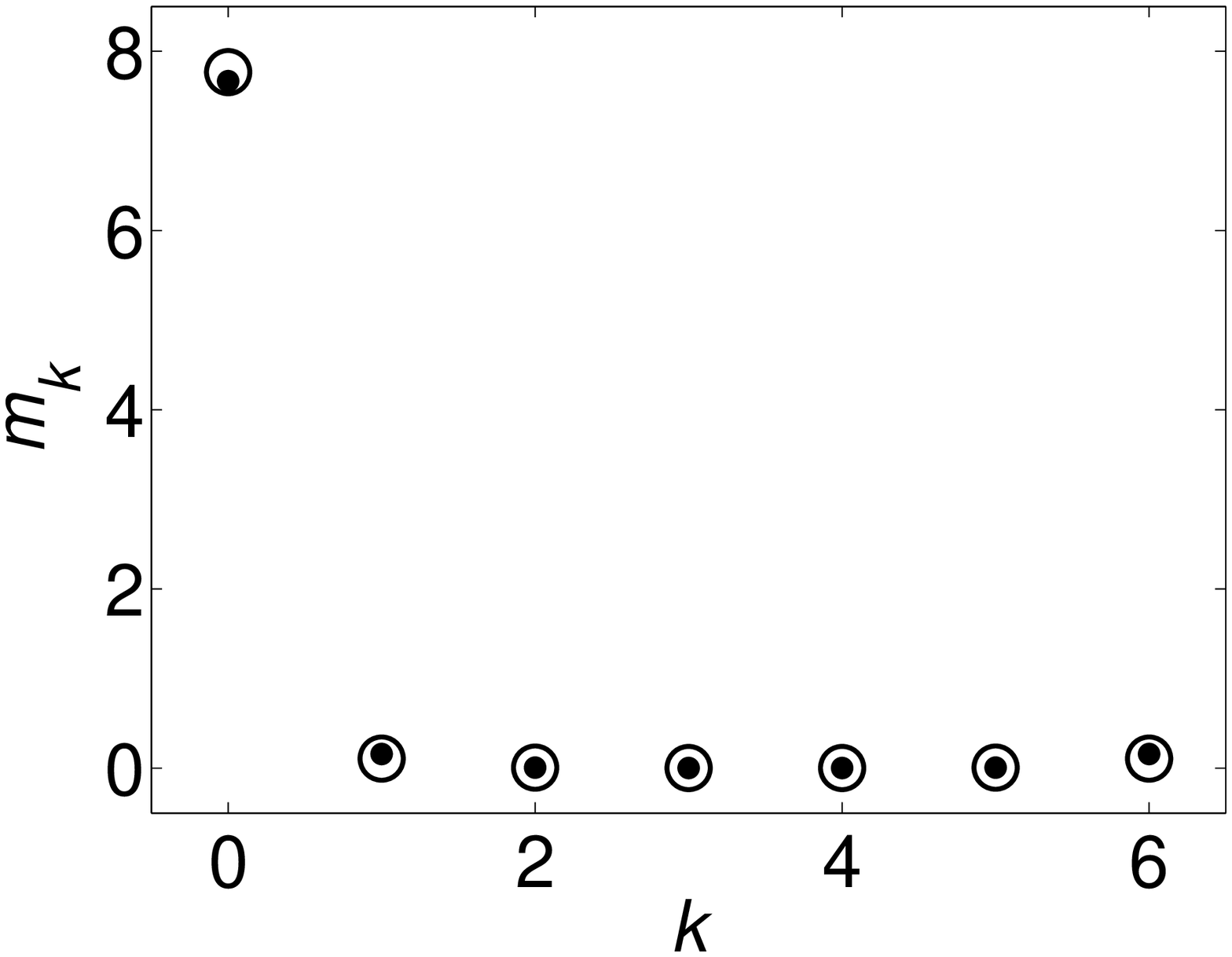}
\includegraphics[width=6.0cm]{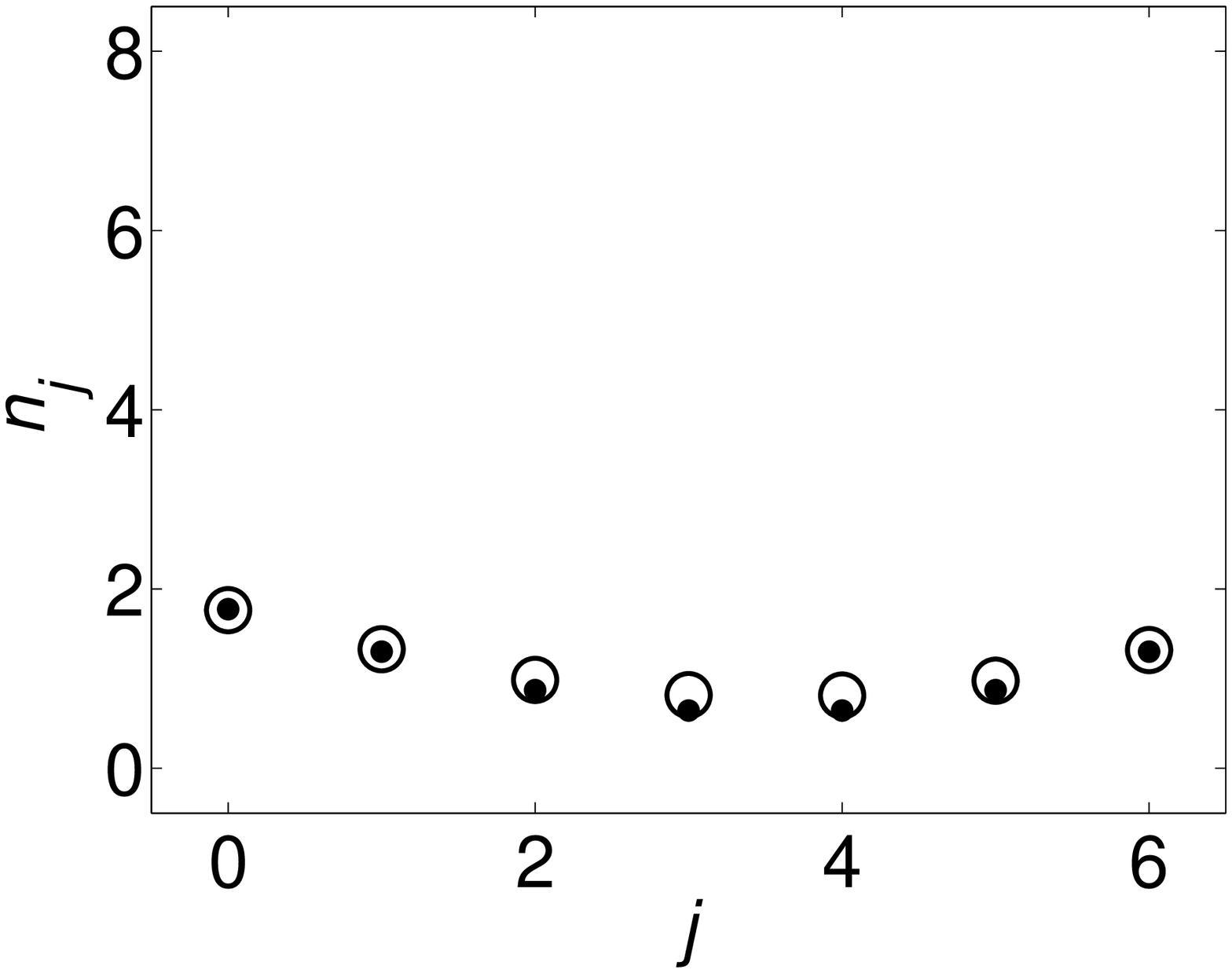}
\end{center}
\caption{
Momentum distribution $m_k$ (left panel) and
space distribution $n_j$ (right panel) of bosons in the ground state for 
$T =\! 1$, $V_0 =\! 0.2$, $U =\! 0.2$, $M =\! 7$, $N = \! 8$. 
Symbols $\bigcirc$ and $\bullet$ have the same meaning as in figure \ref{fig2} and
show an excellent agreement between the distributions relevant to the exact and 
the approximate ground state.
}
\label{fig7}
\end{figure}
%%%%%%%%%%%%%%%%%%%%%%%%%%%%%%%%%%%%%%%%%%%%%%%%%%%%%%%%%%%%%%%%%%%%%%%%%%%%%%%%%%%%%%

\begin{figure}
\begin{center}
\includegraphics[width=6.0cm]{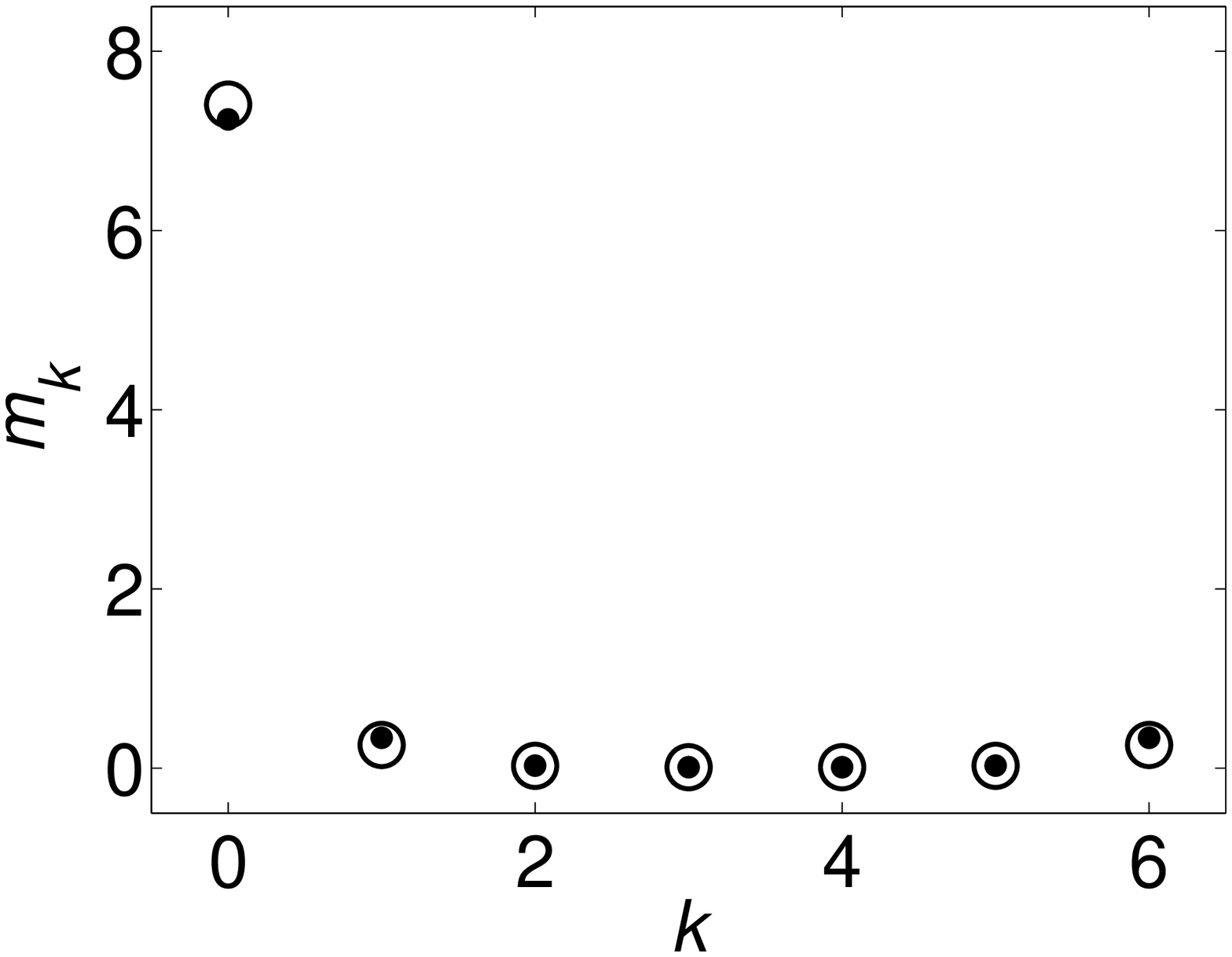}
\includegraphics[width=6.0cm]{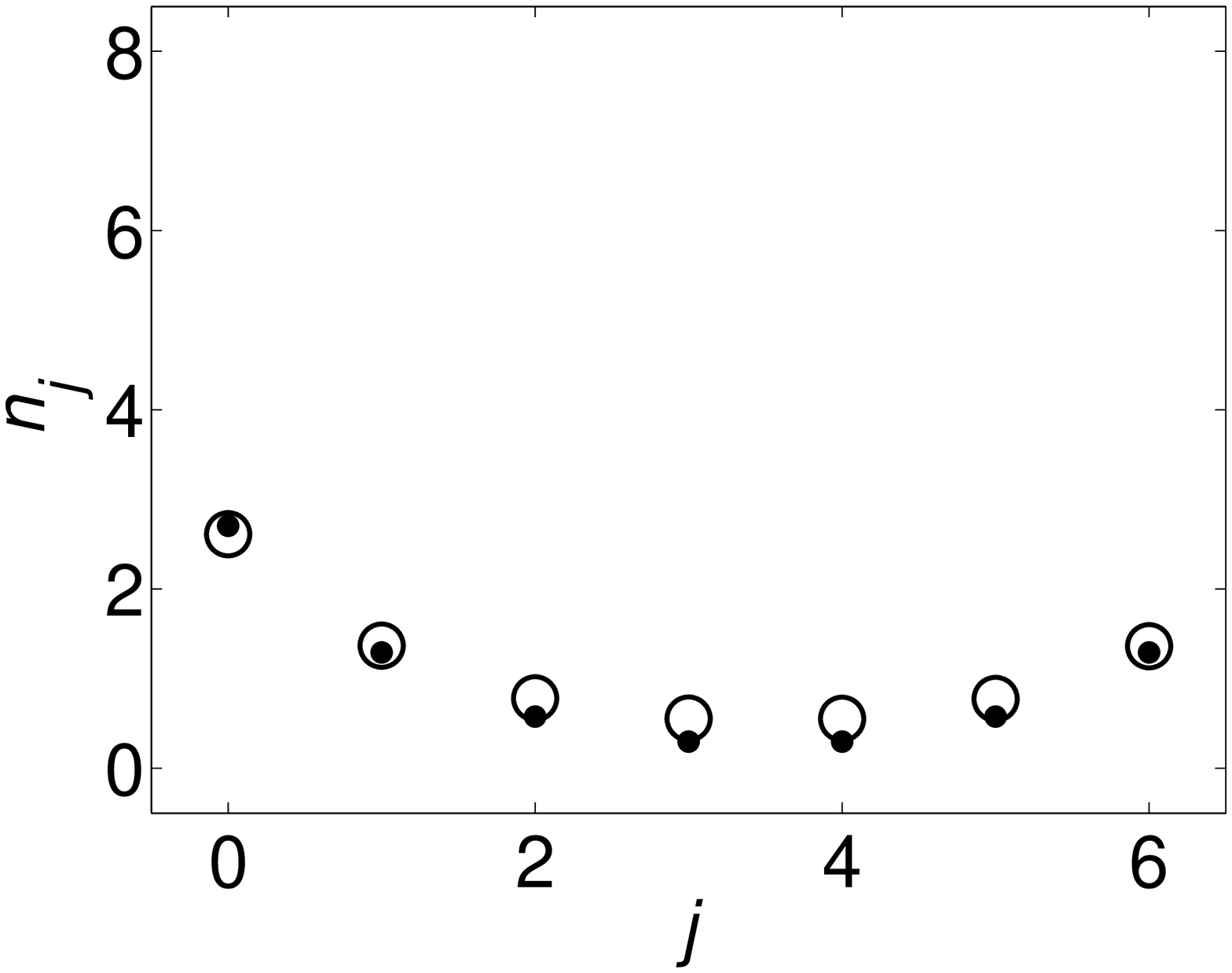}
\end{center}
\caption{
Momentum distribution $m_k$ (left panel) and
space distribution $n_j$ (right panel) of bosons in the ground state for 
$T =\! 1.6$, $V_0 =\! 1$, $U =\! 0.1$, $M =\! 7$, $N = \! 8$. 
Symbols $\bigcirc$ and $\bullet$ have the same meaning as in figure \ref{fig2} and
show an satisfactory agreement between the distributions relevant to the exact and
the approximate ground state.}
\label{fig8}
\end{figure}

As in the case of the SI-regime spectrum, we conclude
by reconstructing the boson distribution both among space modes and among momentum modes
through the formulas $n_i = \langle \mathrm{GS}| a^+_i a_i |\mathrm{GS}\rangle $ and $m_k =\langle \mathrm{GS}| b^+_k b_k |\mathrm{GS}\rangle $
where $| \mathrm{GS} \rangle$ now represents the approximate SF ground state.
Observing that $| \mathrm{GS} \rangle = |E ({\vec p}, {\vec q} )\rangle =RDW \,| {\vec p}, {\vec q} \rangle$ 
with ${\vec p} = 0= {\vec q}$, it is advantageous to define operators
${\cal B}_k = (RDW)^+ b_k RDW$ whose explicit expression is
\begin{equation}
{\cal B}_k = x_k + \frac{1}{2} D_k^+ f_k D_k + 
\frac{1}{2} {\sum}^K_{\ell=1} f_{k \ell} W_\ell^+ C_\ell W_\ell  \, .
\label{fC}
\end{equation}
In this formula $D_k^+ f_k D_k = f_k {\rm ch} (\alpha_k/2) - f^+_k {\rm sh} (\alpha_k/2)$
while $W_k^+ C_k W_k = C_k {\rm ch} (\beta_k/2) + C^+_k {\rm sh} (\beta_k/2)$. The calculation of
$m_k$ remarkably simplifies since $\langle \mathrm{GS}| b^+_k b_k |\mathrm{GS} \rangle = 
\langle  0, 0 | {\cal B}^+_ k{\cal B}_k| 0, 0\rangle$ where the simple ground state 
$| 0, 0 \rangle $ of diagonal Hamiltonian ${\rm H}_f +{\rm H}_C$ can be used.
The resulting momentum mode distribution reads
$$
m_k = \langle b^+_k b_k \rangle = x^2_k + \frac{{\rm sh}^2 (\alpha_k/2)}{2}
+ \frac{1}{2} {\sum}^K_{h=1} f^2_{kh} {\rm sh}^2 (\beta_h /2) 
$$
while the ground-state distribution among space modes 
$n_\ell = \langle a^+_\ell a_\ell \rangle = 
\sum_k \sum_q \exp [ i ({\tilde k} - {\tilde q}) \ell ] \langle b^+_q b_k \rangle/M$
is achieved by resorting again to formula (\ref{fC}) to calculate $ \langle b^+_q b_k \rangle$. 
Figure \ref{fig7} shows that essentially no difference is visible between the values of
$m_k$ and $n_i$ obtained with a ground state determined numerically and 
those supplied by our approximation scheme.
We note that the choice of parameters in figure \ref{fig7} entails that $\tau = 0.625 = 5 v$
and, in particular, $2TM/V_0 = 70 >> 1$. Then the conditions for which solutions $\theta_k \simeq g_k$
in equation (\ref{automu}) are satisfied. In figure \ref{fig8}, where condition $2TM/V_0 >> 1$ is weakened
($2TM/V_0 = 22.4$), distributions based on $| \mathrm{GS} \rangle$ show some deviations from
those based on the exact ground state. Their agreement however is still very satisfactory.
In both cases the approximation $\theta_k \simeq g_k -2V_0/M$ cannot be used since the more restrictive
condition $t>>1$ is never reached.
%
%%%%%%%%%%%%%%%%%%%%%%%%%%%%%%%%%%%%%%%%%%%%%%%%%%%%%%%%%%%%%%%
%%%%%%%%%%%%%%%%%%%%%%%%%%%%%%%%%%%%%%%%%%%%%%%%%%%

%%%%%%%%%%%%%%%%%%%%%%%%%%%%%%%%%%%%%%%%%%%%%%%%%%%%%%%%%%%%%%%%%%%%%%%%%%%%%%%%%%%%%%%%%%
%%%%%%%%%%%%%%%%%%%%%%%%%%%%%%%%%%%%%%%%%%%%%%%%%%%%%%%%%%%%%%%%%%%%%%%%%%%%%%%%%%%%%%%%%%
\section{Conclusions}
\label{conc}

In this paper we have analysed the properties 
%of a system 
of attractive bosons trapped in a 1D optical lattice in the presence of a localized attractive potential.
The two particular regimes that we have considered (the SI and the SF one) allow a completely analytical
approach by means of a Bogoliubov-type approximation.

%Firstly, 
In section \ref{sez2} we observed that in the SI regime the localization of bosons, enhanced
by the presence of a single attractive potential well, allows one to obtain the approximate Hamiltonian
(\ref{HB2}). The diagonalization of the latter gives an excellent description of the main
properties of the ground-state and of the low-excited states, in terms of energy and mean occupation
number in the ambient and momentum space. The comparison between analytical and numerical results
(the latter are necessary to compute the exact spectrum of Hamiltonian (\ref{HV})) 
allows one to identify the region in the parameter space in which the SI hypothesis is satisfied
and thus the approximation is valid.

In addition, the opposite SF regime has been studied in section \ref{sfreg} by applying a
Bogoliubov-type treatment justified by the hypothesis of localization in the momentum space.
The simplification introduced has enabled us to compute the spectrum and the mean occupation number of
particles in the ambient and momentum space. Also in this case, the support of numerical results makes
it possible to show for which values of the parameters the Bogoliubov approximation is actually effective.

The possibility to study our model Hamiltonian in a fully analytical manner is obviously due to the
simple shape of potential $V$ which just reduces to a single localized potential. 
However, despite its simplicity, this model represents an instructive intermediate step toward more structured
systems such as lattices with several defects (local potentials with perturbative depths) or, more in general,
to lattices with many local potentials possibly characterized by random depths:
even if the approach to such systems will be mainly numeric, we expect that
the methodology and the analytical results contained in this paper still represent useful tools.
We expects as well that our approach may be fruitfully applied to other
condensed-matter models. The scheme applied to model (\ref{HbogoV})
in section \ref{sez2}, for example, should be applicable in the diagonalization of the polaron-like
Hamiltonian of mixtures with two atomic species \cite{GPC}. Also, 
simple heuristic calculations show that the repulsive version ($U \to -U$)
of model (\ref{HV}) could be studied through the scheme of section \ref{sez2} for $V_0 > 2T + UN/2$. 
Finally, the knowledge of low-energy states achieved in the attractive case is necessary to investigate
the model when the local potential is time dependent. The study of this case is in progress and
will be discussed elsewhere. 

At the experimental level, the lattice with a single localized potential could be more than a simple
but interesting toy model since it certainly represents the simplest way
to break the translational symmetry of the BH Hamiltonian and thus to make appear the spatial
localization, when present. 
In this respect the realization of toroidal traps \cite{Ryu} with a persistent flow is encouraging.
This system has raised a lot of interest owing to the possibility to create a bosonic Josephson junction
by intersecting the toroidal domain with a transverse laser beam to generate a potential barrier.
In our periodic-lattice model a possible experimental realization of the local potential could
be achieved by using a red detuned laser beam.
%
%To conclude, we underline that the Bogoliubov approach, well-known in the contexts
%of BEC, has been used here in a different way, taking advantage of the large occupation of single-particle
%states in the lattice. 

%%%%%%%%%%%%%%%%%%%%%%%%%%%%%%%%%%%%%%%%%%%%%%%%%%%%%%%%%%%%%%%%%%%%%%%%%%%%%%%%%%%%%%%%%%
%%%%%%%%%%%%%%%%%%%%%%%%%%%%%%%%%%%%%%%%%%%%%%%%%%%%%%%%%%%%%%%%%%%%%%%%%%%%%%%%%%%%%%%%%%
%%%%%%%%%%%%%%%%%%%%%%%%%%%%%%%%%%%%%%%%%%%%%%%%%%%%%%%%%%%%%%%%%%%%%%%%%%%%%%%%%%%%%%%%%%
%%%%%%%%%%%%%%%%%%%%%%%%%%%%%%%%%%%%%%%%%%%%%%%%%%%%%%%%%%%%%%%%%%%%%%%%%%%%%%%%%%%%%%%%%%
\begin{appendix}

%%%%%%%%%%%%%%%%%%%%%%%%%%%%%%%%%%%%%%%%%%%%%%%%%%%%%%%%%%%%%%%%%%%%%%%%%%%%%%%%%%%%%%%%%%%%%%%%%
%%%%%%%%%%%%%%%%%%%%%%%%%%%%%%%%%%%%%%%%%%%%%%%%%%%%%%%%%%%%%%%%%%%%%%%%%%%%%%%%%%%%%%%%%%%%%%%%%

\section{Number of solutions of equation (\ref{SPS})}
\label{A2}

The two parametrizations $\mu = \, {\rm ch} y$ and $\mu =  \,\cos y$, 
allow one to rewrite equation (\ref{SPS}) as
\begin{equation}
%%%\frac{2T M}{w}=  
{\sum}^{M-1}_{k=0}  \frac{1 }{ {\rm ch} y - c_k  }
=
\frac{M \, {\rm sh} (My) }{{\rm sh} y \,  [{\rm ch} (My) -1] }
\, ,
\label{hyperF}
\end{equation}
\begin{equation}
%%%\frac{2T M}{w}= \! 
{\sum}^{M-1}_{k=0}  \frac{1 }{ \cos y - c_k }
= \frac{M \, \sin (My) }{\sin y \,  [\cos (My) -1] }\, ,
\label{trigF}
\end{equation}
giving equations (\ref{SPS1}) and  (\ref{SPS2}). Their extremely simple form is particularly
useful to find single-particle energies $\lambda_k$ both numerically and analytically. 
Nevertheless, the series-like version (\ref{SPS}) of such equations
better elucidates the dependence of the effective number of solutions $\lambda_k$
from parameter $M$. By assuming, in equation (\ref{SPS}),
the equivalent ranges $k\, \in [-\frac{M}{2}+1, \, \frac{M}{2}]$, for $M=2p$, 
and  $k\, \in [-\frac{M-1}{2}, \, \frac{M-1}{2}]$, for $M= \, 2p+1$, one finds
$$
\frac{2T M}{w}=  
\frac{1 }{ \cos y - 1 } +\frac{1 }{ \cos y + 1 }
+{\sum}^{p-1}_{k=1}  \frac{2 }{ \cos y - c_k  }
\, ,
$$
(remind that $c_k = \cos (2\pi k/M)$) for $M= \, 2p$ and
$$
\frac{2T M}{w} \,= 
\frac{1 }{ \cos y - 1}+ 
{\sum}^{p}_{k=1}  \frac{2 }{ \cos y - c_k }
\, ,
$$
for $ M= \, 2p+1$.
It is thus evident how in $\lambda = 2T\, \cos y $ the values of $y$ solving
such equations are in one-to-one correspondence with critical values $y_k= 2\pi k/M > 0$
and that for sufficiently large $TM/w$ one has $y \simeq y_k$. In particular,
for $M=2p$, the number of different eigenvalues is $p = {M}/{2} $, 
while $p = (M-1)/{2}$ is found for $M=2p+1$. 
In both cases term $1 /( \cos y - 1 )$ occurring in the preceding formulas 
does not generate any solution in that it tends to $-\infty$ while ${2T M}/{w}$ 
is positive. Including the isolated solution given by equation (\ref{SPS1}),
the solution number is ${M}/{2} +1$ for $M=2p$ and $(M+1)/{2}$ for $M=2p+1$.

%%%%%%%%%%%%%%%%%%%%%%%%%%%%%%%%%%%%%%%%%%%%%%%%%%%%%%%%%%%%%%%%%%%%%%%%%%%%%%%%%%%%%%%%%%%%%%%%%
%%%%%%%%%%%%%%%%%%%%%%%%%%%%%%%%%%%%%%%%%%%%%%%%%%%%%%%%%%%%%%%%%%%%%%%%%%%%%%%%%%%%%%%%%%%%%%%
\section{Approximate solutions of equation (\ref{SPS2}) }
\label{A3}

For $2T/w >>M/2$, function $z(y) =\frac{2T}{w}\, \sin(y )$ intercepts 
$z(y) = {\rm cot}(My/2) $ in the proximity of its
asymptotes placed at $y =y_k= 2\pi k/M$ in the interval $[0 , \pi]$.
Substituting $y= y_k + \epsilon_k$ 
with $|\epsilon_k| << y_k$ in equation (\ref{SPS2}) gives
$({2T}/{w})\, \sin(y_k+\epsilon_k) = - {\rm ctg} (M \epsilon_k/2)$,
which, with the further assumption $M\epsilon_k <<1$, becomes,
to the second order in $M \epsilon_k$,
$$
\left ( \eta c_k -1 \right )\, \epsilon_k^2 +\eta s_k \, \epsilon_k \,+ {8}/{M^2}\, =0
\, ,\quad
\eta = {8T}/({wM})  \, ,
$$
with $s_k = \sin (2\pi k/M)$ and, as usual, $c_k = \cos (2\pi k/M)$. 
The ensuing solutions are
$$
\epsilon_k = \frac{1}{2(\eta c_k -1)} \left [
- \eta  s_k  \pm {\sqrt { \eta^2 s^2_k - {32} (\eta c_k-1)/{M^2} }}
\right ]
$$
where one should remind that cases $k = 0$ and (in even-$M$ case) $k = M/2$ are excluded.
These solutions are well defined for $\eta >>1$, a condition that perfectly matches 
initial assumption $2T/w >>M/2$. 
For large $\eta$, one obtains $\epsilon_k \simeq -8/(\eta M^2 s_k)$
giving the approximate solutions
$$
{\lambda_k }/{(2T)} = \cos (y_k+\epsilon_k) 
%= \cos ( y_k) \, \cos \epsilon_k - \sin (y_k) \, \sin \epsilon_k
%\simeq \, \cos ( y_k) \, + |\epsilon_k| \, \sin (y_k)
\simeq \, \cos ( y_k) \, + {8 }/{(\eta M^2 ) } 
\, ,
$$
where $\cos (y_k+\epsilon_k) \simeq \cos ( y_k) \, + |\epsilon_k| \, \sin (y_k)$ has been used.
Notice that, at least to the first order in $1/\eta$, energies $\lambda_k$ 
simply represent a shift from values $\cos ( y_k)$.
In order to satisfy condition $\eta = 8T/(wM) >>1$, 
the potential-well depth $w$ (the hopping amplitude $T$) must be much smaller (larger)
than $T$ ($w$). This request, however, is not matched in the SI regime
since $T/UN < 1$ and the effective depth $w = UN +v_0 \simeq UN$ due to $v_0<< UN$.
Then the opposite regime described by the inequality $1 >> 8T/(wM)$ must be considered.

This circumstance suggests to develop a different approximation scheme.
Owing to $1 >> 8T/(wM)$,  function $({2T}/{w}) \sin y$
in equation (\ref{SPS2}) ends up
intercept ${\rm ctg} (My/2)$ close to the zeros thereof. As a consequence $y= {\bar y}_k + \epsilon_k$, 
with ${\bar y}_k = y_k + \pi/M$, where, locally,
${\rm ctg} (My/2) \simeq -{M} (y -{\bar y}_k)/2$. Then,
by considering only first-order terms, equation (\ref{SPS2}) becomes
$({2T}/{w}) \sin ({\bar y}_k + \epsilon_k) \simeq {M}\epsilon_k/{2}$
giving in turn $\epsilon_k \simeq ({4T}/{wM}) \sin ({\bar y}_k)$
and
$$
{\lambda_k }/({2T}) = \cos ( {\bar y}_k + \epsilon_k) \simeq
 \, \cos ( {\bar y}_k) \, +  {4T}\, \sin^2 ({\bar y}_k)/({wM})
\, ,
$$
due to approximation 
$\cos ( {\bar y}_k + \epsilon_k) \simeq \cos ( {\bar y}_k) \, + \epsilon_k \, \sin ({\bar y}_k)$.
%
%%%%%%%%%%%%%%%%%%%%%%%%%%%%%%%%%%%%%%%%%%%%%%%%%%%%%%%%%%%%%%%%%%%%%%%%%%%%%%%%%%%%%%%%%%%%%%%%%%%%%%
%%%%%%%%%%%%%%%%%%%%%%%%%%%%%%%%%%%%%%%%%%%%%%%%%%%%%%%%%%%%%%%%%%%%%%%%%%%%%%%%%%%%%%%%%%%%%%%%%%%%%%

\section{Ground-state boson distribution}
\label{A4}

The boson distrubution in the ambient space involved by state (\ref{GS}) 
is obtained by calculating $\langle \mathrm{GS} | a^+_j a_j| \mathrm{GS}\rangle$. To this end
it is useful to reformulate the ground state in terms of spacelike boson operators.
From equations (\ref{B}) and the fact that  $F^+_0 =b^+_0$,
$F^+_k = (b^+_k + b^+_{-k})/\sqrt 2$ one has
$$
D^+_0 = \sum^K_{k=0} {B}_{0k} \, F^+_k  = 
\sum^K_{k=0} \frac{wr_k}{M} \,\frac{ A(0)}{\lambda_0 - 2Tc_k} \, F^+_k
=
\sum^{M-1}_{k=0} x_k\, b^+_k
$$
with $x_k = { w A(0)}/{[M( \lambda_0 - 2Tc_k)]} $. Then ground state (\ref{GS}) reduces to
$ |\mathrm{GS} \rangle 
%= \frac{ (D^+_0)^{N} }{\sqrt {N !}} \, |0 \rangle
%= \frac{ 1 }{\sqrt {N !}} \, \left (\sum^{M-1}_{k=0} x_k \, b^+_k \right )^{N}\, |0 \rangle\, .
=\left (\sum^{M-1}_{k=0} x_k \, b^+_k \right )^{N}\, |0 \rangle/{\sqrt {N !}} $.
The latter is a SU($M$) coherent state with the normalization condition $\sum_k |x_k|^2 = 1$
(this exactly matches equation (\ref{Ap})) and the properties that 
the momentum-mode and space-mode boson distributions are given by \cite{BP}
$$
\langle \mathrm{GS}| b^+_q b_q |\mathrm{GS} \rangle = N |x_q|^2
\, ,\quad
\langle \mathrm{GS}| a^+_j a_j |\mathrm{GS} \rangle = N |\xi_j|^2
$$
respectively, being 
$$
D^+_0 =\sum^{M-1}_{k=0} x_k\, b^+_k = \sum^{M-1}_{\ell =0} \xi_\ell\, a^+_\ell
\, , \quad
\xi_\ell = \sum^{M-1}_{k=0} x_k\, {e^{i{\tilde k} \ell }}/{\sqrt M} \, ,
$$
owing to definitions (\ref{fab}). After setting $\lambda_0 = 2T {\rm ch} y$, 
the series in $\xi_\ell$ can be computed explicitly giving
%$b_k = \sum_j  a_j\, {e^{-i{\tilde k}j }}/{\sqrt M}$
$$
\xi_j = \frac{w A(0) }{2T M^{3/2}} 
\sum^{M-1}_{k=0}\, \frac{e^{i{\tilde k} j }}{{\rm ch} y - c_k}
=
\frac{w A(0) }{2T M^{3/2} } \, \frac{ {\rm ch} [(M/2-j)y] }{{\rm sh} y \, {\rm sh}(M y/2)}\, .
$$
Equation (\ref{Ap}) provides $A(0)$
$$
\frac{1}{|A(0)|^2} =  
%\frac{ w^2}{M^2} 
\sum^K_{k=0} 
\frac{ w^2 r_k^2 /M^2 }{(\lambda_0 - 2Tc_k)^2}
=
\frac{ w^2}{(2T M)^2} 
\sum^{M-1}_{k=0} \frac{1 }{( {\rm ch} y  - c_k)^2}
$$
$$
=
\frac{ w^2}{(2T M)^2}
\frac{ M + {\rm sh} (M y) \, {\rm coth} y  }{ 2 \, {\rm sh}^2 (M y/2) \, {\rm sh}^2 y}
\, .
$$
Then, the boson distribution in the ambient space reads
$$
\langle \mathrm{GS}| a^+_j a_j |\mathrm{GS} \rangle = N |\xi_j|^2 = 
2N\, \frac{ {\rm ch}^2 [(M/2-j)y] }{ M + {\rm sh} (M y) \, {\rm coth} y  }
$$

%$$ 
%B^p_k = \frac{wr_k}{M} \,\frac{ A(p)}{\lambda_p - 2Tc_k}
%$$
%%%%%%%%%%%%%%%%%%%%%%%%%%%%%%%%%%%%%%%%%%%%%%%%%%%%%%%%%%%%%%%%%%%%%%%%%%%%%%%%%%%%%%%%%%%
%%%%%%%%%%%%%%%%%%%%%%%%%%%%%%%%%%%%%%%%%%%%%%%%%%%%%%%%%%%%%%%%%%%%%%%%%%%%%%%%%%%%%%%%%%%
%
\section{Diagonalization of SF Hamiltonian}
\label{A5}

The action of $R = \prod_{k \ne 0} T_k$ on Hamiltonian (\ref{HbogoV})
entails that $b_k^+ \to b_k^+ + z^*_{k}$ and $b_k \to b_{k} + z_{k}$. The new
Hamiltonian contains a linear term ${\cal L}$ depending on parameters $z^*_{k}$ and $z_{k}$ which
can be removed by exploiting the arbitrariness of $z^*_{k}$ and $z_{k}$.
After some algebra, the new Hamiltonian is found to have the form
$$
{\cal H} = R^+ H R = -\Lambda  + 1/2\, \sum'_{k} g_{k} \bigl (  n_{k} + n_{-k} \bigr )
$$
$$
-1/2\, {\sum}'_{k}   
Un\, \left ( b^+_{-k} b^+_k  +\,  b_{k}  b_{-k} \right ) - V_0 B^+ B + {\cal L} + \Phi
$$
with ${\sum}'_{k} ={\sum}_{k \ne 0}$, where ${\cal L}$ is defined as
$$
{\cal L}
=
{\sum}'_{k} \Bigl [ \, g_{k} \Bigl ( z_{k} b_k^+  + z^*_{k} b_k  \Bigr )
-
Un \left (z^*_{-k} b^+_k  +\,  b_{k}  z_{-k}  \right ) \Bigr ]
$$
$$
 - V_0 \, \sqrt {n} ( B^+ + B )
- \frac{V_0}{\sqrt M} \Bigl ( B {\sum}'_{k} z^*_k + B^+ {\sum}'_{k} z_k \Bigr )  \, ,
$$
and
$$
\Phi = {\sum}'_{k} \frac{ g_{k} }{2} \Bigl (|z_{k}|^2 + |z_{-k}|^2 \Bigr )
-{\sum}'_{k} \frac{Un}{2}  \, \left ( z^*_{-k} z^*_k  +\,  z_{k}  z_{-k} \right )
$$
$$
- ({V_0 }/{M}) {\sum}'_{k} {\sum}'_{h} \, z^*_h z_k - {V_0 }  {\sqrt { n/M }} \, {\sum}'_{k} ( z^*_k +  z_k ) \, .
$$
${\cal L}$ vanishes if the following equations are satisfied
\begin{equation}
V_0 \sqrt {n/M } = g_k z_k - Un z^*_{-k}  - \frac{V_0 }{M} {\sum}_{k \ne 0} z_k \, ,
\label{C1}
\end{equation}
Exploiting the fact that $g_{k} = g_{-k}$ one can show that $z_k \equiv z_{-k} = x_k$.
The new equation for $x_k$'s reads $V_0 \sqrt {n/M} = ( g_k - Un) x_{k}  - X V_0 /{M} $,
with $X =  \sum_{k \ne 0} x_k$, giving
\begin{equation}
x_{k} = \frac{V_0 }{g_k - Un}
\, \left ( \frac{\sqrt {n}}{\sqrt M} + \frac{X }{M}  \right )
= - \frac{V_0 (X + {\sqrt N } ) }{UN -M g_k }\, .
\label{C2}
\end{equation}
Summing on index $k$ on both the left and right-hand side of equation (\ref{C2})
provides $X= -{S {\sqrt N }}/{(1+S)}$ in which $S = \sum_{k \ne 0} {V_0  }/{(UN -M g_k)}$,
and
\begin{equation}
x_k= - \frac{V_0 {\sqrt N } }{(UN -M g_k )(1+S)}
\label{xk}
\end{equation}
determining parameters $x_{k} $. As a consequence, one can simplify the scalar terms $\Phi$ 
depending on $z_k$'s in $\cal H$ finding $\Phi = NV_0 / [ M(1+S) ]$. The Hamiltonian becomes
\begin{equation}
{\cal H} = \! \sum_{k \ne 0} \! \left ( g_{k} n_{k}
-\frac{Un}{2} ( b^+_{-k} b^+_k  +\,  b_{k}  b_{-k}) \right ) - V_0 B^+ \! B -C
\label{qH}
\end{equation}
with $C = \Lambda +\Phi$, and thus
$$
{\cal H} =\!\sum_{h, k \ne 0} \Bigl ( g_{k}\delta_{kh} -\frac{V_0}{M} \Bigr ) b^+_{h} b_k
-\frac{Un}{2} \sum_{k \ne 0} ( b^+_{-k} b^+_k  + \mathrm{H.C.} )  -C  .
$$
The latter can be separated in two independent parts by exploiting operators 
$f_k$ and $F_k$ (see equation (\ref{newop}))
such that $F_0 = b_0$ and $f_0 = 0$ (remind that, if $M$ is even, operator
$F_{M/2} = b_{M/2}$ must be considered while $f_{M/2}= 0$). By observing that 
$
b^+_{-k} b^+_k  + b_{-k} b_k  = (F^+_{k} )^2  + F_{k}^2 -(f^+_{k} )^2  - f_{k}^2 
$
and $n_{k}+n_{-k} = F^+_{k} F_{k} + f^+_{k} f_{k}$,
Hamiltonian ${\cal H}$ reduces to ${\cal H} = {\cal H}_f + {\cal H}_F$
where
\begin{equation}
{\cal H}_f = \! \sum^S_{k=1} \Bigl [ g_{k}  f^+_{k} f_k
+ Un \Bigl ( (f^+_{k} )^2  + f_{k}^2 \Bigl ) \Bigl  ] -C\, ,
\label{qHf}
\end{equation}
\begin{equation}
{\cal H}_F = 
\! \sum^K_{k =1} \Bigl [\, g_{k}  F^+_{k} F_k -Un \Bigl ( F_{k}^2 + \mathrm{H.C.} \Bigr ) \Bigr ] -V_0 B^+ B .
\label{qHF}
\end{equation}
In operator $B = \sum^K_{k = 1} r_k F_k/{\sqrt M}$, apart from $r^2_{M/2} = 1$ when $M$ is even,
$r^2_k = 2$. The ranges of $S$ and $K$ are defined in equations (\ref{Sran}) and (\ref{Kran}),
respectively. 
%
%Remind as well that $K= (M-1)/2$ ($K= M/2$) for odd $M$ (even $M$) and
%$S= (M-1)/2$ ($S= (M-2)/2$) for odd $M$ (even $M$),
%

The third and last step for diagonalizing ${\cal H}$ concerns ${\cal H}_F$ which,
after setting $G_{kh} =g_{k}\delta_{kh} -V_0 {r_kr_h}/{M}$, reads
$$
{\cal H}_F =  {\sum}^K_{k,h =1} G_{kh} F^+_{k} F_h 
-\frac{Un}{2} {\sum}^K_{k =1} \Bigl ( F_{k}^2 + \mathrm{H.C.} \Bigr ) \, .
$$
To get the diagonal form of ${\cal H}_F$ we
define new operators $C_\ell = \sum_h f_{h \ell} F_h$
and $C^+_\ell = \sum_h f_{h \ell} F^+_h$ where $f_{h \ell}$ are undetermined elements of an
orthogonal matrix whose arbitrariness can be exploited to diagonalize matrix $G_{kh}$.
We remind that the orthogonal-matrix properties 
\begin{equation}
{\sum}^K_{h =1} f_{h \ell}f_{h m} = \delta_{\ell m}\, ,\,\,
{\sum}^K_{k =1} f_{\ell k}f_{m k} = \delta_{\ell m}\, ,
\label{feqz}
\end{equation}
are equivalent to the commutation relations
$[C_\ell , C^+_m] = \delta_{\ell m}$ and $[F_\ell , F^+_m] = \delta_{\ell m}$.
Imposing that
$$
{\sum}^K_{k,h =1} G_{kh} F^+_{k} F_h = {\sum}^K_{\ell =1} \theta_{\ell} C^+_{\ell} C_\ell
$$
yields the condition
$\theta_\ell f_{h \ell}  = g_h f_{h \ell} - ({V_0}/{M}) r_h Y_{\ell}$, in which 
$Y_\ell = \sum^K_{k=1} r_k f_{k \ell} $,
giving in turn the two equations 
\begin{equation}
f_{h \ell}  = - \frac{V_0}{M}  \frac{r_h Y_\ell}{ \theta_\ell - g_h }
\, , \,\,
1 = - \frac{V_0}{M} \sum^K_{h =1}  \frac{r^2_h}{\theta_\ell - g_h } \, .
\label{aresul}
\end{equation}
The second equation easily follows from the first one.
% The second equation is obtained by performing the sum $\sum^K_{h =1} r_h f_{h \ell}$
% in the first equation. 
Owing to $g_{h} = g_{-h}$, equation (\ref{aresul}) can be written
in the more general form
\begin{equation}
1 = - \frac{V_0}{M} {\sum}_{h\ne 0} \, \frac{1}{\theta_\ell -g_h } \, .
\label{aresul1}
\end{equation}
Moreover, the calculation of $\sum_h f_{h m}f_{h \ell}$ for $m=\ell$ gives
$|Y_\ell |^{-2} =  ({ V_0^2}/{M^2}) \sum_h \, {r_h^2 }/{(\theta_\ell - g_h )^2}$
thus fixing $Y_\ell$. The final form of Hamiltonian ${\cal H}_F$ is found to be
\begin{equation}
{\cal H}_C = {\sum}^K_{\ell =1} \Bigl [ \theta_{\ell} C^+_{\ell} C_\ell
-\frac{Un}{2} \Bigl ( (C^+_{\ell})^2 + C_{\ell}^2 \Bigr ) \Bigr ]
\label{HFfin}
\end{equation}
thanks to the identity
$\sum_\ell C_{\ell}^2= \sum_{h, k} \sum_\ell  f_{h \ell} f_{k \ell} F_h F_k
=\sum_\ell F_{k}^2$ ($\Rightarrow$ $\sum_\ell (C^+_{\ell})^2 = \sum_\ell (F^+_{\ell})^2$).
This can be easily proven by means of equations (\ref{feqz}). 
\bigskip

%%%%%%%%%%%%%%%%%%%%%%%%%%%%%%%%%%%%%%%%%%%%%%%%%%%%%%%%%%%%%%%%%%%%%
\section{Calculation of parameters $\theta_\ell $}
\label{A6}

Since $g_k = V_0/M +2T (1-c_k)- Un$ equation (\ref{autovtheta})
(equivalent to (\ref{aresul1})) takes the form $ {\cal F} (\mu) = {2T M}/{V_0} $
described by equation (\ref{automu}) where
$$
{\cal F} (\mu) = - \sum_{k\ne 0}  \left [ \, \mu - \left ( 1 - \frac{UN-V_0}{2TM}  - c_k \right ) \right ]^{-1}
%
%\sum_{k\ne 0}  \frac{1 }{ \mu - \left ( 1 - \frac{UN-V_0}{2TM}  - c_k \right ) } = {\cal F} (\mu)
%
%%%%\label{autovth2}
$$
with $\mu \equiv {\theta }/{2T }$.
In the regime ${2T M}/{V_0} >> 1$ one expects that the solutions of such an equation
are values of $\theta$ very close to the asymptote positions $ g_k$. This suggests in turn
to represent $\mu$ as $\mu = V_0/(2TM) - Un/(2T) + 1 -\cos y $ leading to equation
$$
\frac{2T M}{V_0} = \frac{ 1}{ 1 -\cos y  } + {\sum}_{k} \, \frac{ 1}{ \cos y - c_k } \, .
$$
Equation (\ref{autovtheta}) clearly shows how the asymptotes of ${\cal F}( \mu) $
are $K$ and thus one expects to find $K$ solutions.
Thanks to equation (\ref{trigF}), the latter becomes
\begin{equation}
({2T M}/{V_0}) \sin y  \, =  {\rm ctg} (y/2) -M \, {\rm ctg} (My/2)\, .
\label{trigoSF}
\end{equation}
Approximated solutions are found by replacing $y = y_k + \xi_k$ in the latter
formula and using the Taylor expansion to the second order in $\xi_k$. This
supplies the equation
$$
\frac{2T M}{V_0}  (s_k + c_k \xi_k)\xi_k =\frac{ \xi_k s_k - \xi^2_k }{1-c_k} 
-2 +  \frac{M^2}{4} \xi^2_k \, .
$$
($s_k = \sin y_k$, $c_k = \cos y_k$) giving in turn
\begin{equation}
(t c_k -1 + \rho_k ) \xi^2_k  + (t -\rho_k) s_k \xi_k   + {8}/{M^2}  = 0\, .
\label{AtrigoSF}
\end{equation}
with $t = {8T }/{M V_0}$ and $\rho_k ={ 4 }/[M^2 (1-c_k)]$.
At fixed $M$ with $T/V_0$ sufficiently large, $\rho_k$ appears to be negligible with respect
to $t$ for each $k \in [1, K]$. Then the solutions are found to be
\begin{equation}
\xi_k = \frac{1}{2(t c_k -1 + \rho_k )} \Bigl [ - (t -\rho_k) s_k  \pm {\cal R}_k (t, M) \, \Bigr  ]
\label{root1}
\end{equation}
where
$ {\cal R}_k (t, M) \equiv \sqrt { (t -\rho_k)^2 s^2_k - 32 (t c_k -1 + \rho_k )/M^2 }$.

%%%%%%%%%%%%%%%%%%%%%%%%%%%%%%%%%%%%%%%%%%%%%%%%%%%%%%%%%%%%%%%%%%%%%
\end{appendix}
%%%\bigskip

%\bibliography{attractive} 
\end{document}